\documentclass[12pt]{article}

\usepackage{amsmath,amsfonts,amssymb,latexsym,graphicx}
\usepackage{caption}
\usepackage{subcaption}

\newtheorem{theorem}{Theorem}

\newtheorem{definition}{Definition}

\numberwithin{equation}{section}
\def\be{\begin{equation}}
\def\ee{\end{equation}}
\def\bq{\begin{eqnarray}}
\def\eq{\end{eqnarray}}
\def\beq{\begin{eqnarray}}
\def\eeq{\end{eqnarray}}

\begin{document}
\title{\textsc{Dispersive Friedmann universes and  synchronization}}
\author{\Large{\textsc{Spiros Cotsakis$^{1,2}$}\thanks{skot@aegean.gr}}\\
$^{1}$Institute of Gravitation and Cosmology\\ RUDN University,
ul. Miklukho-Maklaya 6, Moscow 117198, Russia\\ \\
$^{2}$Research Laboratory of Geometry,  Dynamical Systems  and Cosmology\\
University of the Aegean, Karlovassi 83200, Samos, Greece\\ \\}
\date{April 2023}
\maketitle
\newpage
\begin{abstract}
\noindent We introduce consideration of dispersive aspects of standard perfect fluid Friedmann cosmology and study  the new qualitative behaviours of cosmological solutions that emerge as the fluid parameter changes and zero eigenvalues appear in the linear part of the Friedmann equations. We find that due to their insufficient degeneracy, the Milne, flat, Einstein-static, and de Sitter  solutions cannot properly bifurcate. However, the dispersive versions of Milne and flat universes contained in the versal unfolding of the standard Friedmann equations possess novel long-term properties not met in their standard counterparts. We apply these results to the horizon problem and show that unlike their hyperbolic versions, the dispersive Milne and flat solutions completely synchronize in the future, hence offering a solution to the homogeneity, isotropy, and causal connectedness puzzles.

\end{abstract}
\newpage
\tableofcontents
\newpage
\section{Introduction}
\subsection{Dispersive dynamics and bifurcations}
It is a mainstream assumption in cosmology that the history of the evolving universe is made up of a number of different epochs such as the vacuum, inflationary, radiation, matter, and other periods, coherently joined to produce the standard cosmological scenario  from some early phase to the future. This scenario usually assumes a cosmic fluid with an equation of state given  by $p=(\gamma-1)\rho$, relating  the fluid density $\rho$  to its pressure through the fluid parameter $\gamma$, specific values of which are associated with special epochs of cosmological evolution, cf. e.g., \cite{cy22},  \cite{weinberg2}.

An efficient way to understand and unify many of the dynamical aspects of a variety of cosmological models is through the application of dynamical systems techniques, cf. e.g., \cite{we}. In this way, using only the simplest themes and ideas from a rich underline dynamical theory, one arrives at a reliable picture of many of the possibilities available in cosmological dynamics, which are  difficult to unravel by other methods.

In introductory dynamical systems treatments (and also in virtually all cosmological applications of the theory to date), one usually deals with a \emph{hyperbolic} flow, namely, one that has only hyperbolic fixed points, i.e., all eigenvalues of  the linearized vector field at each fixed point have nonzero real part (i.e., sinks, saddles, and sources). Standard examples of hyperbolic systems are the  \emph{structurally stable} vector fields which preserve their properties (for example their phase portraits) under small perturbations. All structurally stable systems have hyperbolic fixed points because their linear flows are generic (they form an open and dense set in the space of all linear flows), cf. \cite{hs}, Sec. 16.3.

This  is, however,  a very strong restriction on a vector field. Systems that describe physical phenomena, in particular cosmological dynamical systems, always depend on parameters that are never known exactly, and, in addition, have fixed points that are not hyperbolic, i.e., the linearized jacobian has some eigenvalues on the imaginary axis. In this case, however, contrary to what happens in structurally stable systems, radically new dynamical behaviour may arise.

\emph{Bifurcation theory} is the broad framework that deals with this problem, where one is interested not in the properties of an individual hyperbolic system but in a \emph{family} of dynamical systems that depend on one or more parameters, such that the fixed points are non-hyperbolic for some values of the parameters.  It comprises the study of all sorts of topological changes and qualitative reorganizations of the dynamics that result from a change of the parameters on which the system depends. In distinction to structurally stable systems, no complete analysis exists  of bifurcating or \emph{dispersive} systems. Instead, progress has been made in two fronts: local bifurcation theory which studies  the orbit structure near a non-hyperbolic fixed point of a parametrized  vector field, and global (non-local) bifurcation theory that examines dispersive phenomena which cannot be explained by local bifurcations.

In this case, nearby vector fields can and do have very different orbit structure than that of the individual vector field that corresponds to the value of the parameter associated with the non-hyperbolic fixed point. In an important sense, bifurcation theory is realized by a program of study initiated and emphasized by H. Poincar\'{e} to describe changes in the behaviour of the solutions of an individual system (with non-hyperbolic behaviour) that result from its suitable embedding in a parametric family of dynamical systems,  these changes being unavoidable and unremovable and occur as the parameters vary.

Dispersive methods include bifurcation theory,  but also singularity theory and its applications, the latter  are sometimes called `catastrophe theory' \cite{arny94}. Whereas the theory of bifurcations describes qualitative changes in the phase portraits of dynamical systems, singularity theory is concerned with such changes for more general objects (such as mappings, manifolds, etc) as the parameters change, and because of this it  is very closely related to bifurcation theory. These methods are also  ideal for the description of the emerging behaviours in  complex systems showing some type of collective behaviour, resulting from interactions between their different parts. Such behaviours are again due to the joint effect of two factors, namely, various  degeneracies because of the presence, in the linear part of the vector field, of eigenvalues on the imaginary axis, and secondly because of the presence of parameters.

\subsection{Dispersive problems in cosmology and nonlinear dynamics}
There are various important situations where such dispersive behaviour may arise in a cosmological context. In fact, a fluid parameter, a cosmological constant, the mass of a scalar field, etc, can all be considered as separate parameters in the sense of bifurcation theory. As such, their presence in the cosmological equations is enough to easily produce dynamical systems with an overwhelming dispersive behaviour, leading to novel effects which are totally absent in the standard approach in which these parameters are taken to be constants.

Like in general nonlinear dynamics problems, in cosmology one expects  such phenomena to arise when  a change in a parameter of the system occurs, and will be always accompanied by dynamical behaviour in which the linear part of the vector field has eigenvalues on the imaginary axis corresponding to a `bifurcating value' of the parameter. In such a case, the structure of the linear part of the system determines the degree of `degeneracy', or complexity, of the new equilibria, and leads to possibilities of renewed significance:  the vector field is then expected to `unfold', and the problem becomes one to  capture its extended dynamics.

This approach has to our knowledge not been seriously applied before in a cosmological context, and in this paper we make a start by looking for dispersive behaviour in the simplest, nontrivial cosmological model, a Robertson-Walker spacetime with a perfect fluid with an equation of state $p=(\gamma-1)\rho$, where $\gamma$ is the fluid  \emph{parameter}. This leads to a different situation than that which one is accustomed with in the context of standard Friedmann cosmology, and below we explore the new possibilities to their full extent.
To compare our findings with the standard `hyperbolic' approach to cosmological dynamics, we have also chosen to apply our results to the problem of \emph{synchronization}, a problem that is closely linked  the cosmological horizon problem.

Such `collective' behaviour of \emph{dynamical} synchronization (`sync' hereafter) is in fact very common among different areas of nonlinear dynamics, and eventually wins over possible inherent difficulties in communication \cite{wie58}-\cite{elp}. From cyberspace security and  social networks, to biological and laser oscillators, to chaotically synchronized behaviour, one witnesses an arbitrary number of identical, coupled subsystems gradually approaching  a synced state to remain there forever. The almost magical common property of sync in these systems is that it   emerges spontaneously through  self-organization, in the sense that  although the `receiver' has received only part of the information of the `transmitter', it somehow manages to reconstruct the remaining piece.

In this paper, we consider the evolution of a single Friedmannian domain defined by a Robertson-Walker spacetime satisfying the standard cosmological evolution equations with a single perfect fluid with the equation of state $p=(\gamma-1)\rho$. This represents a single completely homogeneous and isotropic spatial domain evolving in (proper) time. However, the known causality properties of spacetime lead to it containing causal subregions, i.e., parts of any spatial slice of it, that were never in causal contact with each other, yet they now appear to an observer as being completely synchronized - the horizon problem.

The horizon problem of having synchronized but causally disjoint subregions in a homogeneous and isotropic Friedmannian domain, has been known for quite some time, and our description of it completely adheres to the standard one as the following quotations show: In \cite{mtw}, p. 815, we read:

\emph{$\dots$ one concludes that the foundations for the homogeneity
and isotropy of the universe were laid long before the universe became approximately
Friedmann, for if statistical homogeneity and isotropy of the universe had not already
been achieved at the longest wavelengths earlier, these horizon limitations would
have prevented any further synchronization of conditions over large scales while
the universe was in a nearly Friedmann state, and small amplitude ($10\%$) deviations
from isotropy should be observed now$\dots$},

 in \cite{di-pee}, p. 506,

\emph{$\dots$ By comparing radiation background intensities across the sky it is also found that the temperature and expansion rate are precisely synchronized across the visible universe. Even though the separate parts of the visible universe  are not visible to each other they are evolving in precise unison$\dots$}

while in inflation one postulates that such regions were initially `unsynchronized' but were brought to become causally connected very early,  possibly around Planck time,  due to a rapid thermalization process,  \cite{li90}, p. 54:

\emph{$\dots$ expansion began practically simultaneously in different regions of the observable part of the universe with a size $l\leq 10^{-33}\textrm{cm}$, since they all came into being as a result of inflation of a region of the universe no bigger than $10^{-43}\textrm{sec}$ which started simultaneously to within $\Delta t\sim t_P\sim 10^{-43}\textrm{sec}$. The exponential expansion of the universe makes it causally connected at scales many orders of magnitude greater than the horizon size in a hot universe, $R_p\sim ct$$\dots$}.

In addition, a lack of proper causal synchronization of spatial subdomains was generally believed to lead to inhomogeneities and anisotropies, as the following passage from \cite{weinberg1}, pp. 525-6 shows:

\emph{$\dots$ If the homogeneity of the universe is achieved by the physical transport of energy and momentum $\dots$ we should expect the universe $\dots$ [at last scattering] $\dots$ to be inhomogeneous over distances larger than twice the `particle horizon' $\dots$the microwave background ought to exhibit large anisotropies $\dots$ However, there is no sign of any appreciable anisotropy $\dots$ on the contrary, the microwave radiation appears to be highly isotropic [therefore] $\dots$ it is difficult to understand how such a high degree of isotropy could be produced by any physical process occurring at any time since the initial singularity. $\dots$}

In this paper, the possibility  of explaining this problem by a non-instantaneous but gradual, or \emph{dynamical}, synchronization mechanism is proposed as a kind of temporal analogue of  phase transitions\footnote{A totally different kind of synchronization problem based on  Mixmaster domains  all chaotically oscillating and interacting  with weaker couplings and cooperating to sync on approach to the initial singularity, was recently considered in  \cite{ba20,sync1}.}. From the above quotations, we are naturally led to ask the following two questions presently: Can a \emph{single} Friedmann domain that has initially become completely synchronized and homogenized by some process (e.g., inflation)  maintain its synchronization property during its future evolution?  Secondly: can  causally disconnected Friedmann subdomains of the single Friedmann domain synchronize during their evolution even if they were not so initially (so that the single Friedmann domain was also not synced initially)?

Perhaps somewhat surprisingly, in this paper we show that the answer to both questions is `no', if one restricts to the simplest Friedmann-fluid models with hyperbolic equilibria. However, the examination of the dispersive analogues of the simplest Robertson-Walker cosmologies reveals that the models synchronize completely. This we believe provides a new approach to the horizon problem.

\subsection{Outline of this paper}
Our work develops bifurcation phenomena in cosmology, and also points   to a possible connection between bifurcations and sync in a cosmological context.

The structure of this paper is as follows. In the next Section, we provide a more detailed guide to the main results of this paper. Section 3 contains basic  background material in bifurcation theory. This is used in an essential way throughout the remaining of this work. Sections 4-7 contain  results and developments about the dispersive nature of Friedmann cosmologies. In Section 8, we formulate the sync problem in Friedmann cosmology, and then apply our previous results on bifurcating Friedmann universes to this problem. We discuss our findings in the last Section.

\section{A more detailed guide to the main results}
In this Section, we describe the most important results and developments of this work, to guide the reader to the  essential points of our arguments. We may tabulate some essential points and  results as follows:
\begin{enumerate}
\item We compute the center manifolds of all dispersive equilibria of the Friedmann-fluid system.
\item We prove that the unfoldings of the dispersive Milne and flat states are versal.
\item We discuss the implications of the existence of a versal unfolding  for the long-term behaviour of the cosmological solutions.
\item We show that the only systems  that synchronize in the future are the dispersive Milne and flat unfoldings (horizon problem).
\end{enumerate}
A more detailed description of the results of this paper is given below.

Section 3 is devoted to  a summary of various notions and results in basic bifurcation theory. After an introductory  discussion in subsection 3.1, and the two preliminary forms of the basic system (\ref{ds1}), namely, the `linear-Jordan' form,  Eqn. (\ref{jo1}), and the `A-B-C' form, Eqns. (\ref{jo1a})-(\ref{jo1c}) in subsection 3.2,  we present the two fundamental center manifold theorems in Sections 3.3 and 3.4.

These theorems are valid for the non-parametric and parametric center manifolds respectively, and reduce the dimensionality of the original system (\ref{ds1}) to another system of lower dimensionality holding on the center manifold. These theorems also guide  us to compute the center manifold efficiently using the so-called `tangency condition' (parametric or not).

In subsection 3.5, we introduce the notions of bifurcation, codimension, and unfolding. We also discuss the important case where one is able to find one unfolding that includes them all,  the versal unfolding.

In subsection 3.6, we develop these ideas in the simplest context of one-zero-eigenvalue, codimension-1 bifurcations, which correspond to those found in  in later Sections.

Sections 4-7 are the main body of this work. After a preliminary discussion and motivation in subsection 4.1 of how a bifurcation theory approach arises in cosmology for the Friedmann system (\ref{o0})-(\ref{h0}), we present the extra dispersive (i.e., non-hyperbolic) equilibria for single-fluid Friedmann cosmology in Section 4.2. We call these fixed points, type-I, IIa,  IIb-1, and IIb-2, and show that they correspond to dispersive versions of the Milne, flat, Einstein-static, and de Sitter universes respectively.

Then for each one of these basic solutions, we show that the dynamics is suitably reduced to the corresponding parametric center manifold in each case, namely, Eqns. (\ref{ooc}), (\ref{cmEQN1}), (\ref{cmEQN2}), and (\ref{cmEQN3}), respectively (the last case being the most difficult, but perhaps the least interesting one). This results in a huge simplification of the original dynamics given by the system (\ref{o0})-(\ref{h0}).

We then show that the center-manifold-reduced systems violate the necessary conditions (each one for different reasons) and therefore  cannot properly bifurcate. However, the first two, namely the dispersive Milne and  flat equilibria, have versal unfoldings given by Eqns. (\ref{un1a}), (\ref{un1b}) respectively. This result implies that dispersive versions of standard cosmology have  novel cosmological solutions given respectively by Eqns. (\ref{un4}) for the Milne state, and (\ref{un5}) for the flat state, and we discuss their properties.

It is interesting that no such result can hold for the Einstein static and de Sitter spaces in the present context, and we show that these represent  non-generic solutions in this sense (cf. Section 7).

In Section 8, we revisit the two sync questions stated in the Introduction. In subsection 8.1, we show that with respect to sync, the problem  of a distribution of causally disjoint domains is reduced to that of the dynamics of a  single Friedmann domain.

In subsections 8.2, 8.3, we introduce and discuss the sync function and mechanism in general, and for our problem of Friedmann domains  in particular. In the remaining subsections, we show that although the original Friedmann equations cannot usefully synchronize their solutions, their unfoldings where they exist do, and so we conclude that sync in the present context is possible for the dispersive Milne and flat universes.

This then provides an alternative approach to the horizon problem for Friedmann domains, without the need for an initial inflationary stage. In the process of showing the impossibility of sync for the original Friedmann-fluid equations, we perform various explicit flow calculations in subsection 8.4-8.6, but these are possible because the equations are simple in this context.

\section{Center manifolds and bifurcation theory}\label{cm-bif}
In this Section, we describe basic results of bifurcation theory  necessary for the developments in the following Sections. Bifurcation theory originated in   fundamental  work by H.  Poincar\'{e}, was subsequently developed by many mathematicians,  and today forms a central research area in modern dynamical systems. We refer to the basic references \cite{arny83} (abbreviated to [A83]), \cite{guck-ho83} (abbreviated to [GH83]), \cite{wig} (abbreviated to [W3]), and further references therein,  for examples, proofs, and further details. Other valuable works on bifurcations are \cite{arny94}, \cite{arny86}. We assume that the reader has some background in flows, stability, invariant manifolds, and asymptotic sets, as the subject is given in standard sources, e.g., \cite{hs,arny92,ap92}. We exclusively deal below with vector fields, although everything we discuss is directly applicable to maps.

\subsection{Generalities}
We consider the family of  dynamical systems, a `parametrized vector field',
\be \label{ds1}
\dot{Y}=G(Y,\epsilon),\quad Y\in\mathbb{R}^n,\quad \epsilon\in\mathbb{R}^p,
\ee
where $G$ is a $\mathbf{C}^r$ function on some open subset of $\mathbb{R}^n\times\mathbb{R}^p$, and suppose that (\ref{ds1})
has a non-hyperbolic fixed point at $(Y,\epsilon)=(Y_0,\epsilon_0)$. Then the linearized Jacobian $A=D_Y G(Y_0,\epsilon_0)$ (which satisfies the linear system $\dot\xi=A\xi$) has some  eigenvalues on the imaginary axis. The problem we address here is: What is the behaviour of the solutions of  (\ref{ds1}) for $\epsilon$ very close to $\epsilon_0$?

Bifurcation theory  studies all possible ways in which such eigenvalues can arise in the structure of a dynamical system, as well as their  dynamical implications. It classifies the resulting behaviours in terms of the \emph{codimension} of the bifurcation, which is basically the smallest number of parameters that completely describes the bifurcation. The resulting behaviours are described by  the possible \emph{unfoldings}  of the bifurcation, which are families of dynamical systems that contain the bifurcations.

The simplest way in which $(Y_0,\epsilon_0)$ can be non-hyperbolic is when $D_Y G(Y_0,\epsilon_0)$ has \emph{a single zero eigenvalue} and the remaining eigenvalues have nonzero real parts, this belongs to the so-called `codimension-1 bifurcation theory', and is the only case we consider here. (Our treatment does not develop codimension-1 bifurcation  theory with a pure imaginary pair of eigenvalues, nor higher-codimension bifurcations, but in fact also covers the case where  $D_Y G(Y_0,\epsilon_0)$ has \emph{more than one but not all} of its eigenvalues equal to zero and the remaining ones having nonzero real parts.)

Another name for vector fields with some zero or purely imaginary eigenvalues is \emph{dispersive}, borrowing  from similar usage in partial differential equations (cf. \cite{tao}). Hence, we introduce  `dispersive' as another word for `structurally unstable' vector fields.  In this case, a 0-eigenvector defines a \emph{dispersive direction}, while for a nonzero eigenvalue $\lambda$, we call the  $\lambda$-eigendirection with $\lambda$ negative (resp. positive) a \emph{diffusive}, or stable (resp. \emph{anti-diffusive}, or unstable) direction. The set of all bifurcating dynamical systems is a subset of the dispersive ones; in fact, it is a \emph{proper} subset: for example, the equation $\dot{x}=\mu-x^3$ does not bifurcate although it is dispersive. Other examples include certain equations that appear in later Sections of the present paper. In general, dispersive dynamical systems contain new degenerate (i.e., non-generic) behaviours which can only be accounted for by studying  families of systems, not individual ones, cf. [A83], Sect. 29.

\subsection{Normal form of the linearization}
Given the system (\ref{ds1}), there is some preliminary work that needs to be done in order to bring it to a form suitable for subsequent treatment.

We may assume without loss of generality that the fixed point in (\ref{ds1}) is located at $(Y_0,\epsilon_0)=(0,0)$, otherwise we can introduce a linear transformation and transfer it to the origin in both the phase and  parameter space by setting $(v,\mu)=(Y-Y_0,\epsilon-\epsilon_0)$. Then (\ref{ds1}) becomes a system of the form $\dot{v}=H(v,\mu), H(v,\mu)=G(v+ Y_0,\mu+\epsilon_0)$.  We can then split the system  into a linear and a nonlinear part, $\dot{v}=DH(0)v+\bar{H}(v,\mu).$ Using the eigenvector matrix $T$ of $DH(0)$, we can simplify  the latter and write it in Jordan canonical form $J$ under the transformation $v=TX$, so that the full nonlinear system will be written as
\be\label{jo1}
\dot{X}=JX+F(X,\mu),
\ee
where $J=T^{-1}DH(0)T$, and $F(X,\mu)=T^{-1}\bar{H}(TX)$. This is a `normal form' of the system, in which only the linear part $DH(0)$ has been simplified as much as possible.

We shall deal with the case where the Jordan form $J$ that defines  the linear  part of Eq. (\ref{jo1}) is a block matrix of the form,
$
J=\textrm{diag} (A,B,C),
$
where $A$ is a $(c\times c)$-matrix, having eigenvalues with zero real parts, $B$ is a $(s\times s)$-matrix, having eigenvalues with  negative real parts,  and $C$ is a $(u\times u)$-matrix, having eigenvalues with  positive real parts (a basic example is obtained when $A=0,B=-1,C=0$, this is in fact a case that often appears in the present work). Then the system (\ref{jo1}) becomes,
\beq
\dot{x}&=& Ax+f(x,y,z,\mu)\label{jo1a}\\
\dot{y}&=& By+g(x,y,z,\mu)\label{jo1b}\\
\dot{z}&=& Cz+h(x,y,z,\mu)\label{jo1c},
\eeq
with $f,g,h$ denoting the nonlinear part of the vector field. Under our assumptions, $\bar{H},F,f,g,h$ are all $C^r$ functions of their respective coordinates.

Before we proceed further, we note an important fact about the way  we deal with \emph{parametric} systems (cf. [A83, p. 269], [W3, Sec. 18.2]). This is to regard the parameter  $\mu$ as one of the \emph{dependent} variables by adding the trivial equation,
\be\label{mu}
\dot{\mu}=0,
\ee
in the system (\ref{jo1a})-(\ref{jo1c}) as an extra  equation. This, however,  increases the number of dispersive (i.e., zero real part) eigenvalues from $c$ to $c+p$, where $p$ is the number of parameters (components of the vector $\mu$). In addition, we shall regard terms of the form $\mu_i x_j,\mu_i y_j,\mu_i z_j$ (with the indices $i,j$ in their appropriate ranges each time) as \emph{nonlinear} terms, so that they will  not be present in the linearized Jacobian.

The original system (\ref{ds1}) is now in  suitable form in order to examine how bifurcation considerations enter in the study of parametrized systems.

\subsection{The center manifold}
Let us first consider the case where there are no unstable directions, that is $C=0, h=0$ in Eq. (\ref{jo1c}), and also no parameters. In this case, the system (\ref{jo1}) instead of having the form (\ref{jo1a})-(\ref{jo1c}),  becomes,
\beq
\dot{x}&=& Ax+f(x,y)\label{jo11a}\\
\dot{y}&=& By+g(x,y)\label{jo11b}
\eeq
where $f,g$ are $C^r$ functions, and $A,B$ as before.
We assume  that this system has a fixed point,
\be
f(0,0)=0,\quad g(0,0)=0,
\ee
which is dispersive (that is non-hyperbolic),
\be
Df(0,0)=0,\quad Dg(0,0)=0.
\ee
We give the following definition.

\begin{definition}[Center manifold]
A center manifold is an invariant manifold for the system (\ref{jo11a})-(\ref{jo11b}) that can be locally represented as the graph,
\be
W^c_{loc}=\left\{(x,y)\in\mathbb{R}^2|\,y=h(x),|x|<\delta, h(0)=0, Dh(0)=0 \right\},
\ee
for sufficiently small $\delta$.
\end{definition}
We note that here $x$ denotes the dispersive dimension, while $y$ is the diffusive (stable) dimension of the system. (We also use the letter $h$ to denote the center manifold function, without a danger of confusing it with the same letter we used before for the nonlinear function associated with the unstable dimensions.)

The fundamental  properties of (non-parametric) center manifolds are contained in the following theorem.
\begin{theorem}\label{cm0}
For the system (\ref{jo11a})-(\ref{jo11b}) there exists a $C^r$ center manifold having the following properties:
\begin{enumerate}
\item \textbf{Reduction:} The dynamics of the system on the center manifold is given by the following $c$-dimensional `reduced' system ($c$ is the number of zero eigenvalues),
\be
\dot{x}= Ax+f(x,h(x)).
\ee
\item \textbf{Stability:} The  stability of the center manifold solution $x(t)$ is the same as the stability of the full solution $x=x(t),y=h(x(t))$ of the original system, and, for sufficiently small $x$, the two solutions (that captured by the center manifold and the original one)  are identical up to exponentially small terms.
\item \textbf{Approximation:} The center manifold can be computed to any finite order of $x$ for small solutions by using the tangency condition, $\dot{y}-Dh(x)\dot{x}=0$.
\end{enumerate}
\end{theorem}
We refer to \cite{carr81} for a proof.

\subsection{Center manifolds with parameters}\label{par-method}
The theory developed in the previous subsection has a simple but important extension when we include the parameter $\mu\in\mathbb{R}^p$. In this case, instead of the system (\ref{jo11a})-(\ref{jo11b}) we have the parametric system,
\beq
\dot{x}&=& Ax+f(x,y,\mu)\label{jo12a}\\
\dot{y}&=& By+g(x,y,\mu)\label{jo12b}\\
\dot{\mu}&=&0\label{jo12c}.
\eeq
We assume that this system has a dispersive fixed point at the origin,
\be\label{dis}
f(0,0,0)=0,\quad g(0,0,0)=0, \quad Df(0,0,0)=0,\quad Dg(0,0,0)=0,
\ee
and that similar assumptions hold  as before for  $A,B,f,g$. Then Theorem \ref{cm0} still holds on the \emph{parametric center manifold},
\be\label{par-graph}
W^c_{loc}=\left\{(x,\mu,y)\in\mathbb{R}^3|\,y=h(x,\mu),|x|<\delta_1, |\mu|<\delta_2, h(0,0)=0, Dh(0,0)=0 \right\},
\ee
for the \emph{reduced vector field},
\beq
\dot{x}&=& Ax+f(x,h(x,\mu),\mu)\label{jo13a}\\
\dot{\mu}&=&0\label{jo13c},
\eeq
that is, we have,
\begin{theorem}\label{cmp}
For the system (\ref{jo12a})-(\ref{jo12c}) there exists a $C^r$ parametric center manifold given by (\ref{par-graph}), such that the reduced dynamics on $W^c_{loc}$ is given by (\ref{jo13a}). The stability properties of the solutions of (\ref{jo12a})-(\ref{jo12c}) are described by those of the reduced vector field (\ref{jo13a})-(\ref{jo13c}) up to exponentially small terms.  The parametric center manifold can be computed to any finite order of $x$ for small solutions by using the tangency condition, $\dot{y}-D_x h(x,\mu)\dot{x}=0$.
\end{theorem}
We refer to [A83], [GH83], and [W3] for further comments and for the original references.

We note that for the two center manifold theorems above, the inclusion of the remaining unstable (anti-diffusive) dimensions for both the parametric and the non-parametric cases, is direct, but of course the fixed point would then be unstable. However, all  this behaviour will again be captured (or, `realized') on the center manifold (cf. [A83], pp. 268-270, [W3], Sec. 18.3), and for this reason the full system (\ref{jo1a})-(\ref{jo1c}), (\ref{mu}) is sometimes called \emph{a suspension over the center manifold family (\ref{jo13a})-(\ref{jo13c})} (cf. [A83] for this terminology).

Although center manifolds of fixed points  are non-unique, their Taylor expansions agree to all orders (cf. [W3], p. 264). In addition, center manifolds are tangent to the center eigenspaces at the dispersive fixed points, and also depend smoothly on the parameter $\mu$. Therefore center manifolds present a huge simplification for studying the behaviour of a dispersive system because the full dynamics (as described by the suspension) gets  constrained (when `projected') on them and becomes independent of the choice of the center manifold (for more details, see [A83], Sect. 32C).

\subsection{Degeneracy, codimensions, and unfoldings}\label{de-co-un}
In this subsection we provide a general, intuitive discussion of basic notions of bifurcation theory.

We shall say that (the dynamics of) two (possibly parametric) vector fields $f,g$ (or simply their flows) are \emph{$C^k$-equivalent} ($k=0,1,2, \dots$) if there is a $C^k$-map that takes the orbits of the flow of $f$ to the orbits of the flow of $g$ preserving their orientation. If in addition the map preserves parametrization by time, then we say that  $f,g$ are \emph{$C^k$-conjugate.}  When the  flows of  $f,g$ are $C^k$-equivalent,  the fixed points of $f$ are mapped to the fixed points of $g$ and the same is true for their  periodic orbits, but their periods may not be equal. However,  $C^k$-conjugate flows have the same mapped eigenvalues and equal periods of their mapped periodic orbits. For instance, the famous Hartman-Grobman theorem says that every hyperbolic flow is $C^0$-conjugate to its linearization (cf. e.g., [W3], p. 350).

We now give the following basic definition. We consider a  system of the form (\ref{jo13a})-(\ref{jo13a}), namely,
\beq
\dot{x}&=& F(x,\mu)\label{gena}\\
\dot{\mu}&=&0\label{genb},
\eeq
that is a general system after center manifold reduction. We shall also assume that (\ref{gena}) is one-dimensional.
\begin{definition}[Bifurcation]\label{def-bif}
We say that the fixed point at the origin, $(x,\mu)=(0,0)$, of the system (\ref{gena})-(\ref{genb}) \emph{undergoes a bifurcation at $\mu=0$} if the flow of $F(x,\mu)$  for $\mu$ near zero and $x$ near zero is \emph{not} $C^0$-equivalent to the flow of $F$ at $\mu=0$  and $x$ near zero, that is of $F(x,0)$. In this case we say that the point $(x_0,\mu_0)=(0,0)$ is a \emph{bifurcation point} of the system (\ref{gena})-(\ref{genb}).
\end{definition}

As we remarked above, the condition  that the fixed point is non-hyperbolic is a necessary but not a sufficient one for it to be a bifurcation point. Intuitively speaking, the nature of a bifurcation of the system is such that when the parameter $\mu$ is varied through the bifurcation point, the system (i.e., the vector field) \emph{itself} is `deformed' and new dynamical phenomena occur, for example,  fixed points can be created or annihilated.

The most distinct  characteristic of a bifurcation is the existence of one or more \emph{bifurcation sets}, that is smooth submanifolds of fixed points in the phase-parameter space $(x,\mu)$ of the form,
\be\label{bifcurve}
\mu=\mu(x).
\ee
Bifurcation sets, in addition to the fixed point condition $F(x,\mu(x))=0$ (as implied by the implicit function theorem), also satisfy certain \emph{non-degeneracy} and \emph{transversality} conditions.

The simplest example of such sets are the \emph{bifurcation curves} in one-dimensional, one-parameter (i.e., $p=1$)  systems.
These curves are most easily pictured by drawing them in a \emph{bifurcation diagram}, which for a system of the form (\ref{gena})-(\ref{genb}) is represented in the $(\mu-x)$-plane. The system finds itself `moving' on these bifurcation curves, and their properties completely describe the nature and new dynamics associated with the ensuing bifurcation as the parameters of the system are changed.

More generally, the goal is to study how the solutions of the equation $F(x,\mu)=0$ change as the parameter is varied, e.g., for what $\mu$-values solutions appear or disappear? Such questions are studied in detail in the sister fields of bifurcation theory, \emph{singularity theory}, and \emph{catastrophe theory}, and their consideration becomes very essential for bifurcation problems  \cite{arny94,arny86,gg73,arny85,arny2012}.

For a given parametric family of systems such as (\ref{gena}), as the parameter of the system varies through its bifurcation value and the vector field changes, we say that  the system `bifurcates', or `deforms', or `unfolds', and the problem becomes one to classify all these different `unfoldings', that is all those families which contain the given  bifurcation in a `persistent' way (cf. [GH83], p. 123). As the parameter changes, the unfoldings contain those behaviours of the system which are unremovable and so should not be missed. This is so because the removal of some degeneracy at one individual value of the parameter in one unfolding, appears anew at a nearby value of another  (one cannot remove all degeneracies simultaneously, e.g., \cite{arny86}, p. 16).

To study the unfoldings of a given family, we must perturb it in suitable ways. Not all non-hyperbolic fixed points are equally degenerate, and their degree of degeneracy  is measured by the degree of instability they exhibit under such perturbations. The least degenerate bifurcations are those that are stable, that is those families that retain their bifurcation features when perturbed, whereas others have perturbations that result in new bifurcations.

It turns out that the most useful invariant to classify  the possible behaviour of parametric systems under parameter changes is the \emph{codimension of the bifurcation}. This  equals the smallest number of parameters present in the new system into which we need to embed the family  in order to capture all possible qualitative dynamics that can occur near the bifurcation point. If this number equals to $k$, then we speak of a codimension-$k$ bifurcation. A complete study of systems with $k>2$ is generally an important open problem in modern mathematics.

In some cases, the study of all possible unfoldings of a system can be obtained from a single one that yields all other possible bifurcations. This is called the \emph{versal deformation}, or the \emph{universal unfolding} of the system, and was first studied by Poincar\'{e} in his Thesis. Bifurcation theory provides  general methods to compute the codimension of a bifurcations and, in particular, to construct  the \emph{versal unfolding} (we use this term collectively to describe both versal deformations and universal unfoldings (although there are generally subtle differences between the two concepts, these are discussed in \cite{gs85,gss}).

\subsection{The simplest bifurcations with a single-zero eigenvalue}
The remaining problem is to describe the dynamics of (\ref{ds1}) in the special case  of the reduced family (\ref{jo13a})-(\ref{jo13c}) on the center manifold. We shall assume that $A=0$ (but ($B\neq0$), this way we have a 1-dimensional system with a single zero eigenvalue on the center manifold and the remaining eigenvalues  have nonzero real parts\footnote{Here we do not discuss the Hopf bifurcation, the more complicated case where there is a pair of imaginary eigenvalues instead of a single-zero eigenvalue. However, we note the interesting fact that dynamical considerations  restrict the codimension of the Hopf bifurcation (and also others) to a smaller value than that found by using singularity theory considerations  alone (1 instead of 2). This explains the word `simplest' in the title of this subsection.}.

The problem is therefore reduced to describing the dynamics of the 1-dimensional system (similar to the general system (\ref{gena})-(\ref{genb})),
\be\label{one}
\dot{x}=f(x,\mu),\quad x\in\mathbb{R},\quad \mu\in\mathbb{R}^p,
\ee
near the origin, when,
\be
f(0,0)=0,\quad Df(0,0)=0,
\ee
so that it has a non-hyperbolic fixed point at the origin (in both phase and parameter space) with a single-zero eigenvalue, and no terms  linear in x (we note that a term of the form $\mu x$ is nonlinear of order 2). We shall also assume that $p=1$ (when more than one parameters are present we regard all of them except one as fixed).

The simplest bifurcations of the system (\ref{one}) are given below. In each case, we give the name of the bifurcation and the `normal form' of the vector field (the function $f(x,\mu)$):
\begin{enumerate}
\item The normal form of the \emph{saddle-node bifurcation:} \be\label{sn}\dot{x}=\mu-x^2.\ee
\item The normal form of the \emph{transcritical bifurcation:} \be\label{tc}\dot{x}=\mu x-x^2.\ee
\item The normal form of the \emph{pitchfork bifurcation:} \be\label{pf}\dot{x}=\mu x-x^3.\ee
\end{enumerate}
There are \emph{bifurcation conditions} on $f$ for the  behaviour of the system (\ref{one}) to be one of these prototypical  three types on the center manifold, but we shall not be repeated them here (see,  [A83], Sect. 32, [GH83], p. 148, Thm.  3.4.1, [W3], Sect. 20.1). These conditions ensure that in each type there are  bifurcation curves that describe the nature of the bifurcations.

An important property of the three bifurcations above is their distinctive behaviour under perturbations.  While the saddle-node bifurcation is stable under perturbations in the sense that the addition of perturbation terms does not introduce new phenomena, the transcritical bifurcation is unstable and breaks into either a pair of curves of fixed points without bifurcating, or into a pair of saddle-node bifurcations, and the same is true for the pitchfork family (see [W3], Sect. 20.3). We say that in this case  the \emph{degeneracy} of the fixed point increases from the saddle-node, to the transcritical, to the pitchfork bifurcations.

The term `normal form of the saddle-node bifurcation' introduced above means that the dynamics of \emph{any} system of the form (\ref{one}) with $f(x,\mu)=a_0\mu+a_1x^2+a_2\mu x+a_3\mu^2$ and having a saddle-node bifurcation is qualitatively the same as that of the system $\dot{x}=\mu\pm x^2$. All $O(3)$-terms or higher could be neglected without qualitatively  affecting the dynamics.

In fact, one may show that the saddle-node and the transcritical are codimension-1, but the pitchfork is a codimension-2 bifurcation. In terms of versal (unfolding) families, one may better understand the significance, or typicality, of these bifurcations, in particular that of the saddle-node's:  The system $\dot{x}=ax^2, a:$ \textrm{const}, has the saddle-node bifurcation $\dot{x}=\mu+ax^2$ as its versal unfolding, whereas the fixed point of  $\dot{x}=ax^3, a:$ \textrm{const} is of codimension-2, with versal unfolding the pitchfork bifurcation $\dot{x}=\mu_1 +\mu_2 x+ax^3$ (cf. [W3], Example 20.4.10,  for a proof of these results).
Thus the saddle-node is the `generic' codimension-1 bifurcation (cf. [A83], pp. 267-8, for a somewhat alternative description of this result in terms of the so-called  `$\check{S}$o$\check{s}$itai$\check{s}$vili's reduction').

Finally, we note that besides the bifurcation diagram as an efficient way of depicting the normal forms of the 1-dimensional bifurcations, there are two alternative but very illuminating descriptions of the dynamics of these bifurcations:
\begin{enumerate}
\item The \emph{`metamorphoses' of the phase portraits} given by the $(x,y)$-planes for $\mu<,=,$ or $>0$, and
\item The \emph{phase-parameter portraits} given by the  $(x,y,\mu)$ 3-space.
\end{enumerate}
For the phase portrait metamorphoses of all three bifurcations discussed above (in the case of one diffusive dimension), we refer the reader to the \cite{perko}, p. 340-1; also in [GH83], p. 126, there is the phase-parameter portrait of the (full suspension of) the pitchfork bifurcation. It is an instructive exercise to draw the corresponding two- and  three-dimensional portraits for all remaining cases of possible suspensions for the three basic codimension-1 bifurcations.

\section{Dispersive Friedmann domains}
In this Section, after some remarks mainly for motivating the idea of cosmological bifurcations, we discuss the dispersive fixed points of the Friedmann fluid system.

\subsection{A bifurcation approach to Friedmannian evolution}
As discussed in the Introduction, we consider  a single Friedmannian causal domain $\mathcal{A}$, that is a spatial region with all `points' (or subregions inside it) `homogenized'. (For the moment we shall restrict our attention only to this case, leaving to subsection 8.1 the more general situation where we also  allow for  subdomains of $\mathcal{A}$ to be causally disconnected with respect to others.)

The evolution of $\mathcal{A}$ is governed by the usual dynamical equations (cf. \cite{we}, Sec. 2.3), for the Hubble parameter $H=\dot{a}/a$, defined using the scale factor $a$ as a function of the proper time $t$ (a dot is derivative with respect to $t$), and  the density parameter  $\Omega=\rho/(3H^2)$, with  the fluid density $\rho$ related to the pressure by $p=(\gamma-1)\rho$, $\gamma$ being the fluid parameter. Then using a dimensionless time variable $\tau$  defined by $dt/d\tau=1/H$,  the evolution  equations for $\mathcal{A}$ are given by,
\begin{align}
\frac{d\Omega}{d\tau}&=-\mu\Omega+\mu\Omega^2\label{o0}\\
\frac{dH}{d\tau}&=-H-\frac{1}{2}\mu\Omega H.\label{h0}
\end{align}
Here  $q=\mu\Omega/2$ denotes the deceleration parameter, and we have set $\mu=3\gamma-2$.

Below we shall be interested in the question of whether or not the system (\ref{o0})-(\ref{h0}) is dispersive, and, in particular, whether or not it admits a bifurcation set of the form (\ref{bifcurve}). However,
keeping $\mu$ constant,  as is the case in the usual approach to cosmological dynamics  described by (\ref{o0})-(\ref{h0}), cannot provide a suitable framework to address this question, because in this way one would  restrict the evolution of Friedmann domain  $\mathcal{A}$  to a particular era in its evolution.

We shall assume instead that the (\ref{o0})-(\ref{h0}) defines a \emph{family} of differential equations
parametrized by  $\mu$ which is now promoted to be a true continuous \emph{parameter} of the problem\footnote{In the standard model of cosmology, evolutionary properties  of $\mathcal{A}$ are really taken to always depend only on some given fixed value(s) of $\mu$, in other words, the ratio $`w=p/\rho$' is constant.  This allows us for instance, to consider `mixtures' of different, non-interacting fluids, and calculate the present pressure, and other quantities of interest in this case, cf. \cite{weinberg2}, Sect. 1.5. Here,  instead, the emphasis is different, we allow for \emph{passages} through different eras corresponding to different values of $\mu$, such that radiation, cold dark matter, baryonic matter, etc., to play a role. This is similar to studying  the dynamics of a pendulum when the properties of the system are allowed to depend on two continuous parameters, its length and the strength of gravity; this problem never arises when taking a `mixture' of different pendula of given unequal lengths and gravity strengths.}. This is a crucial point that plays an important role below, and we shall be interested in the properties of the solutions of the system (\ref{o0}-(\ref{h0}) \emph{as $\mu$ is varied}. This leads us to consider issues that lie beyond  the `hyperbolic' behaviour met in the usual approach to cosmological  dynamics, and only necessarily emerge as we enter the realm of bifurcation theory.

But what is the physical significance of adopting $\mu$ as a continuous  \emph{varying parameter} rather than a constant as is usually the case in cosmology? We shall now provide some extra motivation and examples that illustrate this important distinction, with the hope to  make this point clearer. This is also particularly timely, especially for gravitational studies, where it is not unusual for various such `constants' to appear  in the governing systems of equations.

In order to understand the various possibilities, it is useful to distinguish between  two broad categories of  dynamical systems: \emph{generic} and \emph{degenerate}.  For the former, it was shown by Poincar\'{e} that the only behaviour of the phase curves that occur in a neighborhood of an equilibrium is either a focus, or a node, or a saddle. This behaviour associated with generic systems is also called \emph{hyperbolic}, to distinguish it from more complicated ones, like say a (non-hyperbolic) center, or periodic orbit,  as in the case of the simple harmonic oscillator, $(y,-\omega_0^2 x)$ for each frequency, at the origin. This degenerate behaviour as is well-known  is, however, unstable with respect to small generic perturbations of the vector field, such as $(y,-\omega_0^2 x -\epsilon y)$, for $\epsilon$ constant: the center is altered into a (hyperbolic)  sink or a source and is consequently destroyed into stable (for $\epsilon>0$) or unstable (for $\epsilon<0$) foci.  That is, more complicated cases turn into generic ones under small generic perturbations of the system.

However, this situation changes dramatically if we are interested not in an individual system but in a \emph{family} of systems that depend on one or more varying parameters. For concreteness, let us consider \emph{the space of all dynamical systems}, $S$, and imagine it divided into regions of generic ones separated by dividing walls of degenerate systems. Under a small change of the parameters,  a degenerate system belonging to one of the walls, becomes generic. For instance, let us suppose that   a curve  in $S$ representing a 1-parameter family of systems intersects a wall at a non-zero angle (i.e., transversally). The point of that intersection corresponds to a particular value of the parameter defining the family of systems (represented here by a curve). We now consider a small perturbation of this curve which intersects the wall at some different nearly point. Now the intersection point of the original curve with the wall has move to another point on the new curve \emph{off that wall}, and so has become non-degenerate because it now belongs to the domain of generic systems (off that wall). However - and this is the crucial point here - the new curve still intersects the wall at a new point, that is for a different value of the parameter, so a new degeneracy arises there. Hence, the degeneracy that appeared for some parameter value and eliminated by a small perturbation, now appears anew at another value of the parameter. We conclude that \emph{when  a family of systems with a continuous varying parameter rather than an individual system is considered, degenerate cases are not removable}. This situation becomes more pronounced when the number of parameters is increased, for 1-parameter systems only the simplest, the codimension-1, degeneracies appear. Hence,  if we artificially restrict the parameters of a dynamical system to be constants instead of true, continuously varying objects, all degenerate cases are essentially lost, because they are removed from consideration when the system is generically perturbed. Thus we can be hugely misled in our study of dynamical systems in this way. We now give two examples that illustrate this situation.

Let us first consider the example of the Duffing's equation, $\ddot{x}+\delta\dot{x}-\beta x + x^3=\gamma\cos\omega t$, a nonlinear oscillator exhibiting spectacular phenomena. This  has, for $\gamma=0,\delta>0$, and $\beta$  assumed a fixed constant, a sink for $\beta<0$, and two sinks and a saddle for $\beta>0$, cf. [GH83], pp. 84-5. Should one stop the analysis at this point, and conclude that the system can exhibit only these hyperbolic equilibria? Is this a complete analysis of the possible behaviours of the system? Certainly not! Keeping $\gamma=0,$ the system undergoes a pitchfork bifurcation as $\beta$ passes through zero if considered as a continuous parameter, the linear part of the system evaluated at the origin has eigenvalues $0$ and $-\delta$, cf. [GH83], pp. 134. In this case the stability of the system is altered completely, and depends on its reduction to the center manifold and the associated bifurcation diagram (cf. [GH83], pp. 136). Further, when we consider the parameter  $\gamma\neq 0$, period doubling phenomena, attractors, and braided structures appear, with the braids repeating on finer and finer scales as the attractor folds indefinitely, as is common also with other systems having chaotic attracting sets. As the parameter $\gamma$ increases, the attractor disappears and the story given by the bifurcation diagram becomes more complicated. All this novel behaviours are completely absent if we restrict our analysis only to the hyperbolic case.

Let us give one last example to illustrate the distinction between constants vs. continuous varying parameters. This is a Lorenz dynamical system, a simplified model for fluid convection in a 2-dimensional layer heated from below. This system is described by the vector field $(\sigma(y-x),\rho x+x-y-xz,-\beta z+xy)$, with $(x,y,z)\in\mathbb{R}^3, \sigma, \beta$ positive constants, and $\rho$ a parameter (usually it is typical to put $\rho=r-1$ in the standard version of the Lorenz equations). Below we shall consider only the possibility where two of the parameters are fixed and only one remains a true varying parameter (the $\rho$).  If for the moment we consider $\rho$ also as a positive \emph{constant}, then it is easy to see that the system has a hyperbolic fixed point at the origin, a saddle (when $\rho>0$), and further  hyperbolic equilibria which are kinds of spiral saddles (one positive and two complex eigenvalues with nonzero real parts). For $\rho<0$, the origin is a hyperbolic sink, and becomes the unique attractor of the system. This is the overall hyperbolic behaviour of the Lorenz system. However, as is well known, the Lorenz system contains much more structure than the simplified hyperbolic analysis suggests, even in the simple case we consider, with only one of the parameters as varying ($\dot{\rho}=0$). Indeed, for $\rho=0$, a pitchfork bifurcation occurs like in the unforced Duffing oscillator, and further bifurcations happen for higher values, denoted here by $\rho_h$, of $\rho$ (cf. [GH83], chap. 3, and refs. therein). These include a \emph{subcritical} Hopf bifurcation which shows that the standard solutions become unstable and are replaced by new large amplitude motions as the parameter $\rho$ passes through $\rho_h$. In fact, the Lorenz attractor exists for a wide range of parameter values $\rho$, while when $\rho$ is sufficiently high, a strange attractor has been rigorously shown to exist as a result of the ensuing period doubling bifurcations.

Other examples of such behaviour abound, and can be found  by consulting the references quoted in the reference list at the end of this paper.

We now return to our problem.
Our approach in this paper is to search for possible \emph{bifurcation} properties of the dynamical solutions for $(\Omega,H,\mu)$ with respect to  the varying parameter $\mu$. In fact, in its most standard version, bifurcation theory employs variation of parameters \emph{without}  any further  dependence of the parameters on the time or on any other of the variables\footnote{Although something like that could be done, it would hugely complicate matters, which are already far remote from any usual situation met in the hyperbolic case.}. This is also what is followed below: we assume that $\mu$ is a new variable of the problem that  satisfies the equation $d\mu/dt=0$,   eventually added to the system (\ref{o0})-(\ref{h0}) as a separate equation.

\subsection{Dispersive Friedmann equilibria}
In sharp contrast to the usual approach to cosmological dynamics but in line with what has been discussed above, we shall take the system (\ref{o0})-(\ref{h0}) to play the same role as that which the primary system (\ref{ds1}) played for subsequent developments in  Section 3,  and look for the  possible existence of dispersive fixed points.

Hence,  we set $Y=(\Omega,H)$, $G(Y,\mu)=(-\mu\Omega+\mu\Omega^2,-H-\frac{1}{2}\mu\Omega H)$, and  $\epsilon=\mu$, presently. Then  the Jacobian $J_{Y}G$ for (\ref{o0})-(\ref{h0}) is,
\be J_{(\Omega,H)}G=\left(
  \begin{array}{cc}\label{jac0}
    -\mu +2\mu\Omega& 0 \\
    -\frac{1}{2}\mu H & -1  -\frac{1}{2}\mu\Omega\\
  \end{array}
\right).
\ee
There are five types  of equilibria  $(Y,\mu)=(Y_0,\mu_0)$, solutions of the equation,
\be\label{G}
G(\Omega,H,\mu)=(0,0),
\ee
for the system (\ref{o0}-(\ref{h0}):
\begin{enumerate}
\item \textbf{EQ-H1:} Hyperbolic Milne  states: $(\Omega,H,\mu)=(0,0,\mu_0)$, for any constant $\mu_0$.
\item \textbf{EQ-H2:} Hyperbolic flat states: $(\Omega,H,\mu)=(1,0,\mu_0)$, for any constant $\mu_0$.
\item \textbf{EQ-I:} Dispersive type I states: The  $\Omega$-axis, $(\Omega,H,\mu)=(\Omega,0,0).$
\item \textbf{EQ-IIa:} Dispersive type IIa states: The flat state $(\Omega,H,\mu)=(1,0,0)$.
\item \textbf{EQ-IIb:} Dispersive type IIb states: The $H$-axis $(\Omega,H,\mu)=(1,H,-2)$.
\end{enumerate}
\textbf{EQ-H1, EQ-H2} are hyperbolic for any $\mu_0$ and are well understood.

For each of the remaining fixed points in the list, namely,  \textbf{EQ-I}, \textbf{EQ-IIa}, \textbf{EQ-IIb}, the corresponding Jacobian  (\ref{jac0}) has a zero eigenvalue. These fixed points only come about provided we regard $\mu$ not a constant but as a continuous varying parameter. This important distinction, although perhaps new to cosmology, is in fact very common and in full compliance with the principles of bifurcation theory as discussed earlier. It is also  a standard procedure followed in dynamics in other parts of physics as discussed in the previous subsection, so that the `constant vs. parameter' distinction for  $\mu$ in this paper becomes completely  analogous to  that  distinction in those examples.

In our problem, the  equilibria \textbf{EQ-I}, \textbf{EQ-IIa}, \textbf{EQ-IIb} have the following characters:

\textbf{EQ-I} is dispersive: Setting  $\mu=0$, we need $H=0$ in (\ref{G}) in order to have an equilibrium solution; in particular, in this case we have a line of fixed points of the form $(\Omega,0,0)$, for any $\Omega$. Then the  Jacobian (\ref{jac0}) becomes $J_{\textrm{EQ-I}}=\textrm{diag}(0,-1)$ in this case. We shall consider the origin $(0,0,0)$, and we may call this a `dispersive Milne' state.

\textbf{EQ-IIa} is a dispersive version of the  flat  state: For $\Omega=1$, if we set $\mu=0$, we can have the  fixed point solution $(1,0,0)$ from Eq. (\ref{G}). The Jacobian from Eq. (\ref{jac0}) is then, $J_{\textrm{EQ-IIa}}=\textrm{diag}(0,-1)$. (We consider this case separately from \textbf{EQ-I} because cosmologically it corresponds to a different solution of the Friedmann equations, namely the flat rather than the Milne solution,  although as we shall see the dynamical treatments of both are very similar.)

\textbf{EQ-IIb} is another dispersive choice:  setting again $\Omega=1$, if we require $\mu=-2$ then we have an equilibrium  for any $H$, the line of fixed points  $(1,H,-2)$. The Jacobian then becomes, $J_{\textrm{EQ-IIb}}=\,\textrm{diag}(-2,0)$. This case, as we shall see later, contains the dispersive versions of the Einstein static universe and of de Sitter space.

The behaviour of the corresponding solutions near the equilibria \textbf{EQ-I}, \textbf{EQ-IIa}, \textbf{EQ-IIb} is  analyzed qualitatively below, as orbits in the corresponding phase spaces. We shall refer to them as  \emph{Type-I, IIa, and IIb solutions}. We emphasize that these solutions correspond to non-hyperbolic (i.e., dispersive) behaviour, and consequently their properties cannot be deduced from the linearized parts of the systems, the same way as  in hyperbolic problems.

The main tool for a complete analysis of  these issues is bifurcation theory, and in this work we shall use this tool to obtain a detailed picture of the dynamics. Once this is done, we shall see that the system (\ref{o0})-(\ref{h0}), albeit being non-hyperbolic, cannot properly bifurcate.

In our ensuing  investigations  to explain this remarkable conclusion for the Friedmann system (\ref{o0})-(\ref{h0}), we shall discover two novel  situations: Firstly, that the failure to bifurcate lies in a deeper effect, that  of violation of the fundamental bifurcating conditions. Secondly, we shall be led to the existence and construction of a unique perturbation of the Friedmann equations (\ref{o0})-(\ref{h0}) that has the sought for bifurcation property. This in turn will have interesting implications for the structure of the universe.

\section{Dispersive type-I states}

\subsection{The center manifold}

\subsubsection{Preparation of the equations}

Near \textbf{EQ-I}, we can write the system (\ref{o0})-(\ref{h0}) in the form (\ref{jo1}), namely, $X'=JX+F(X,\mu),$ where $X=(\Omega,H)^\top$, and,
\be
\left(
  \begin{array}{c}
    \Omega' \\
    H' \\
  \end{array}
\right)
=
\left(
  \begin{array}{cc}
    0 & 0 \\
    0 & -1 \\
  \end{array}
\right) \left(
  \begin{array}{c}
    \Omega \\
    H \\
  \end{array}
\right)
+
\left(
  \begin{array}{c}
    -\mu\Omega+\mu\Omega^2 \\
    -\frac{1}{2}\mu\Omega H \\
  \end{array}
\right),
\ee
with `\,$'\,$' standing for `$d/d\tau$'. This is of the form (\ref{jo12a})-(\ref{jo12c}), with $A=0, B=-1$, namely,
\begin{align}
\Omega'&=-\mu\Omega+\mu\Omega^2\label{sys1}\\
H'&=-H-\frac{1}{2}\mu\Omega H\label{sys2}\\
\mu'&=0,\label{sys3}
\end{align}
the nonlinear part of the vector field  being,
\be\label{nonlin}
F(\Omega,H,\mu)=(f(\Omega,H,\mu),g(\Omega,H,\mu))=\left(-\mu\Omega+\mu\Omega^2,-\frac{1}{2}\mu\Omega H\right).
\ee
In the  form (\ref{sys1})-(\ref{sys3}),  the original system (\ref{o0})-(\ref{h0}) is  a parametric system with parameter $\mu$ in the sense of Section \ref{cm-bif}.

Using Eq. (\ref{nonlin}), the Milne state \textbf{EQ-I} is dispersive:  since,
\be
Df=(-\mu+2\mu\Omega,0),\quad Dg=-\frac{1}2{}\mu(H,\Omega),
\ee
 the conditions (\ref{dis}) are satisfied, namely,
\be
f(0,0,0)=0,\quad g(0,0,0)=0, \quad Df(0,0,0)=0,\quad Dg(0,0,0)=0.
\ee
As we saw earlier,
the Jacobian of the system (\ref{sys1})-(\ref{sys3}) at $(\Omega,H;\mu)=(0,0;0)$ is,
\be J_{(0,0)}=\left(
  \begin{array}{cc}\label{j1}
    0 & 0 \\
    0 & -1 \\
  \end{array}
\right),
\ee
with eigenvalues  $0,-1$. The center eigenspace is then the set $H=0$, that is the $\Omega$-dimension, while the diffusive eigenspace (i.e., that corresponding to the eigenvalue $-1$) is the set $\Omega=0$, that is the $H$-axis. Both sets $\Omega=0, H=0$ are invariant.

\subsubsection{Center manifold reduction}

It follows from the parametric center manifold theorem (Theorem 2 of Section \ref{par-method}) that for the system (\ref{sys1})-(\ref{sys3}) a center manifold exists for  $\mu$ sufficiently close to $\mu=0$, given by,
\be\label{cenman}
W^c_{loc}=\left\{(\Omega,\mu,H)\in\mathbb{R}^3|\,H=h(\Omega,\mu),|\Omega|<\delta_1,|\mu|<\delta_2, h(0,0)=0, Dh(0,0)=0 \right\},
\ee
with $\delta_1,\delta_2$ sufficiently small. This  represents a graph over the $\Omega,\mu$ variables,  $H=h(\Omega,\mu)$, while
all  solutions of the system (\ref{sys1})-(\ref{sys3})  satisfy the conditions of the center manifold reduction.

The next thing is to explicitly find the dynamics on the center manifold. An efficient  way to do this is to compute the function  $h(\Omega,\mu)$ by direct calculation using the tangency condition of Section \ref{par-method}. Taking the time derivative of the center manifold function  $H=h(\Omega,\mu)$, $\dot{H}=D_\Omega h\,\dot{\Omega}$,  using  (\ref{o0})-(\ref{h0}), and substituting,
\be
h(\Omega,\mu)=a\Omega^2+b\mu \Omega+c\mu^2+O(3),
\ee
the tangency condition reads,
\be
D_\Omega h\, (-\mu\Omega+\mu\Omega^2)+h+\frac{1}{2}\mu\Omega h=0.
\ee
Using the parametric center manifold theorem, we can find an approximate solution to any desired degree of accuracy. Equating terms of like powers, and neglecting terms of orders $O(\Omega^2\mu), O(\mu^3),\dots$, we find that $a=b=c=0$.

The result is  that the center manifold $W^c_{loc}$ from Eq. (\ref{cenman}) is the $\Omega$-axis, and the reduced dynamics is governed by,
\begin{align}
\Omega'&=-\mu\Omega+\mu\Omega^2\label{ooc}\\
\mu'&=0.\label{m0}
\end{align}
We note the important fact that in Eq. (\ref{ooc}), quadratic terms include the $\Omega^2$ term. The significance of this will be seen more clearly later. Since  the other eigenvalue is negative, it follows from the center manifold theorem that the unstable manifold $W^u_{loc}$ is empty,  and hence, all solutions of the full system are \emph{stably} attracted by the center manifold (i.e., rapidly decay to it).

\subsection{Do dispersive Friedmann type-I models bifurcate?}

\begin{figure}
     \centering
     \begin{subfigure}[b]{0.3\textwidth}
         \includegraphics[width=\textwidth]{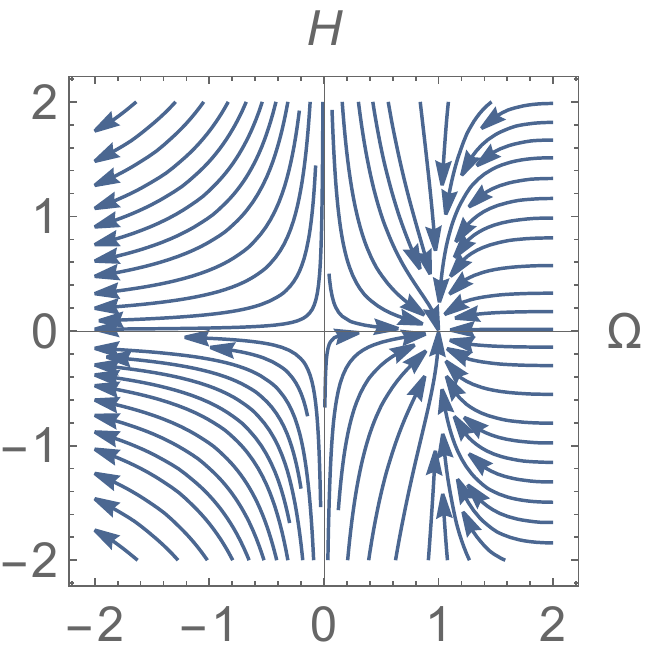}
         \caption{$\mu=-1$}
         \label{mu-1}
     \end{subfigure}
     \hfill
     \begin{subfigure}[b]{0.3\textwidth}
         \includegraphics[width=\textwidth]{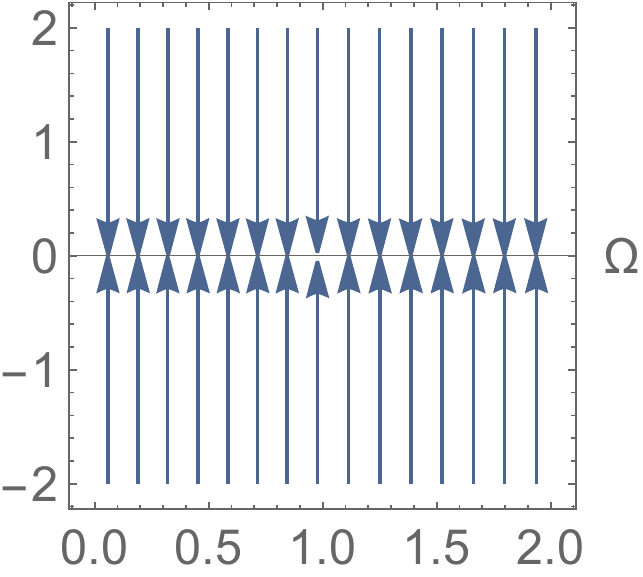}
         \caption{$\mu=0$}
         \label{mu-0}
     \end{subfigure}
     \hfill
     \begin{subfigure}[b]{0.3\textwidth}
         \includegraphics[width=\textwidth]{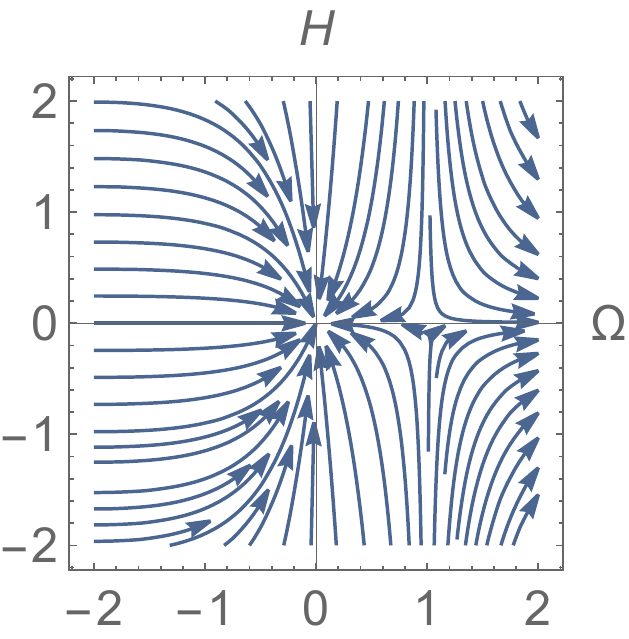}
         \caption{$\mu=+1$}
         \label{mu-+1}
     \end{subfigure}
             \caption{Evolution of  Friedmannian domain $\mathcal{A}$ according to the system (\ref{sys1})-(\ref{sys3}) at and near the parameter value $\mu=0$, showing the behaviour  near the fixed point \textbf{EQ-I}.}
        \label{m-fig}
\end{figure}

In Fig. \ref{m-fig}, we see three examples of  phase portraits for the family of systems in  (\ref{sys1})-(\ref{sys3}), for $\mu=0$ and two values of $\mu$ near zero. Since the diagrams for $\mu\neq 0$ have no zero eigenvalue, they are qualitative different than that at $\mu=0$, and one may conclude that the system \emph{apparently} satisfies the general conditions of a bifurcation in the Definition \ref{def-bif} of Section \ref{de-co-un}.

In practice, however, we need the existence of \emph{a curve of fixed points} for any system to undergo a bifurcation at the bifurcating  value $\mu=0$\footnote{As we shall see below, this will illustrate the somewhat  subtle property that the non-hyperbolicity of a fixed point is a necessary but \emph{not} a sufficient condition to warrant a proper bifurcation.}.
We can use the center manifold reduction theorem of Section \ref{par-method} and focus on the reduced vector field  (\ref{ooc}) to examine more carefully the question of whether the field bifurcates and, in particular, how this relates to the  codimension-1 bifurcations (\ref{sn})-(\ref{pf}).

\begin{figure}
\begin{center}
\includegraphics[scale=0.9]{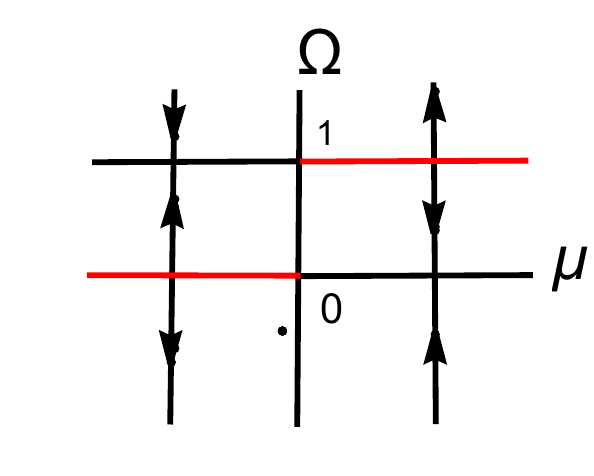}
\end{center}
\caption{Phase-parameter diagram of the system (\ref{ooc})-(\ref{m0}). The  red  lines show instability and the horizontal black ones signify stability. The $\Omega$ phase lines are the lines orthogonal to the $\mu$-axis.}
\label{b1}%
\end{figure}

We can see the behaviour of $f(\Omega,\mu)$ more clearly, if we draw the phase-parameter diagram for the vector field (\ref{ooc}), as in Fig. \ref{b1}.

The global evolution of the system is characterized by the different signs of the $\mu$-parameter: when $\mu<0$, the equilibrium $(0,0)$ is unstable and $(1,0)$ is stable, when $\mu>0$, $(0,0)$ is stable and $(1,0)$ is unstable, while at $\mu=0$ we have the fixed point \textbf{EQ-I}.

However, apart from the appearance of the hyperbolic fixed points \textbf{EQ1} and \textbf{EQ2} for $\mu\neq 0$, there is \emph{no} relation of the form $\mu=\mu(\Omega)$ that could lead to a bifurcation curve, that is no new curve of fixed points passing through the origin to create some kind of bifurcating behaviour.

This conclusion is also clearly seen in the corresponding phase diagrams, for different $\mu$-values for the system (\ref{o0})-(\ref{h0}) in  Fig. \ref{m-fig}. We only see saddle and nodes throughout the evolution for $\mu\neq 0$, but no `saddle-node' exists created at the origin, as one would expect by drawing the corresponding phase portraits of the normal forms of the codimension-1 bifurcations (\ref{sn})-(\ref{pf}).
\begin{figure}\label{fig-tot1}
     \begin{subfigure}[b]{0.3\textwidth}
         \includegraphics[width=\textwidth]{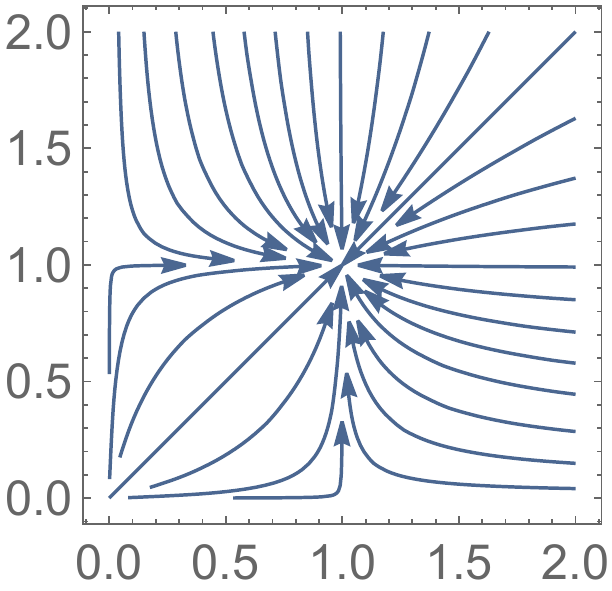}
     \end{subfigure}
      \hfill
     \begin{subfigure}[b]{0.3\textwidth}
         \includegraphics[width=\textwidth]{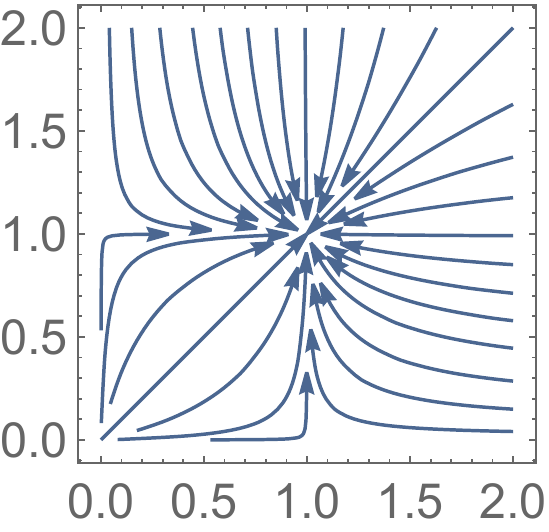}
     \end{subfigure}
     \hfill
     \begin{subfigure}[b]{0.3\textwidth}
         \includegraphics[width=\textwidth]{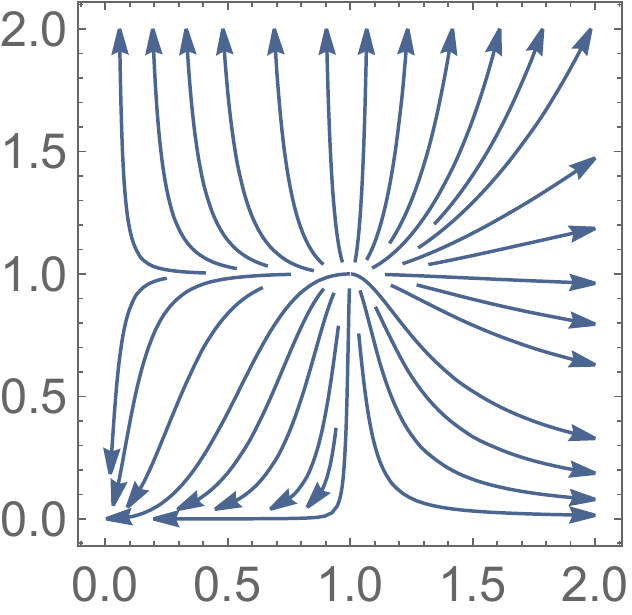}
     \end{subfigure}
\caption{($\Omega_1,\Omega_2$)-phase portraits for $\mu_1$, $\mu_2$ in the cases of: (a) vacuum-vacuum, (b) radiation-radiation, and (c) dust-radiation.}
 \end{figure}

The same conclusion can be  observed from the center manifold reductions of different $\mu$-values. In Figures 3-5, we have plotted the $\Omega_1,\Omega_2$ reduced phase lines, for a number of  different pairs of values $\mu_1, \mu_2$, which satisfy equations of the form (\ref{ooc}). We again only observe saddle and nodes, but no `connecting', or bifurcating, behaviour.

 \begin{figure}\label{fig-tot2}
      \begin{subfigure}[b]{0.3\textwidth}
         \includegraphics[width=\textwidth]{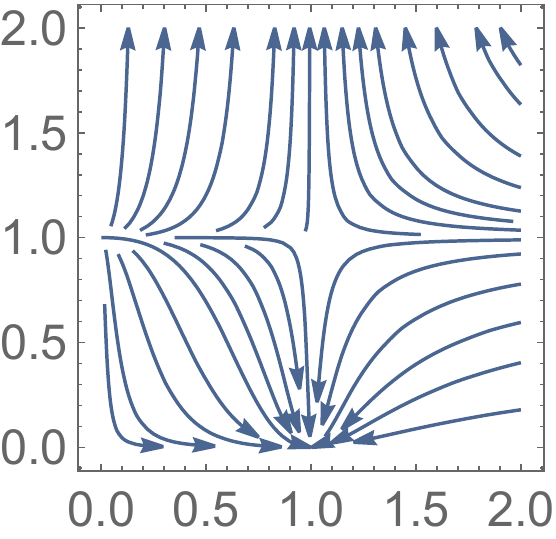}
     \end{subfigure}
     \hfill
     \begin{subfigure}[b]{0.3\textwidth}
         \includegraphics[width=\textwidth]{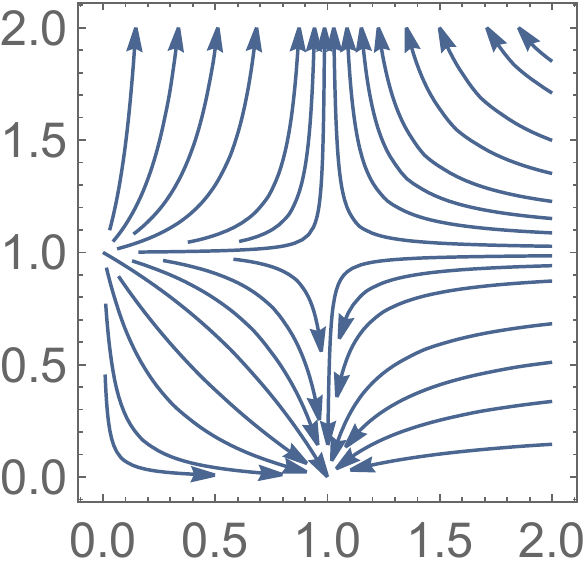}
     \end{subfigure}
       \hfill
     \begin{subfigure}[b]{0.3\textwidth}
         \includegraphics[width=\textwidth]{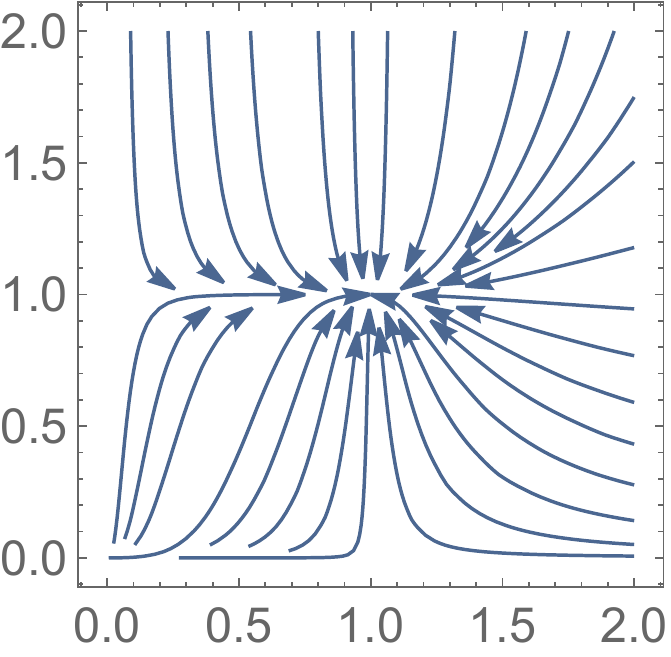}
     \end{subfigure}
\caption{($\Omega_1,\Omega_2$)-phase portraits for $\mu_1$, $\mu_2$ in the cases of: (a) vacuum-scalar field, (b) vacuum-radiation, and (c) vacuum-($\gamma=-1$).}
 \end{figure}

 \begin{figure}\label{fig-tot3}
     \centering
     \begin{subfigure}[b]{0.3\textwidth}
         \centering
         \includegraphics[width=\textwidth]{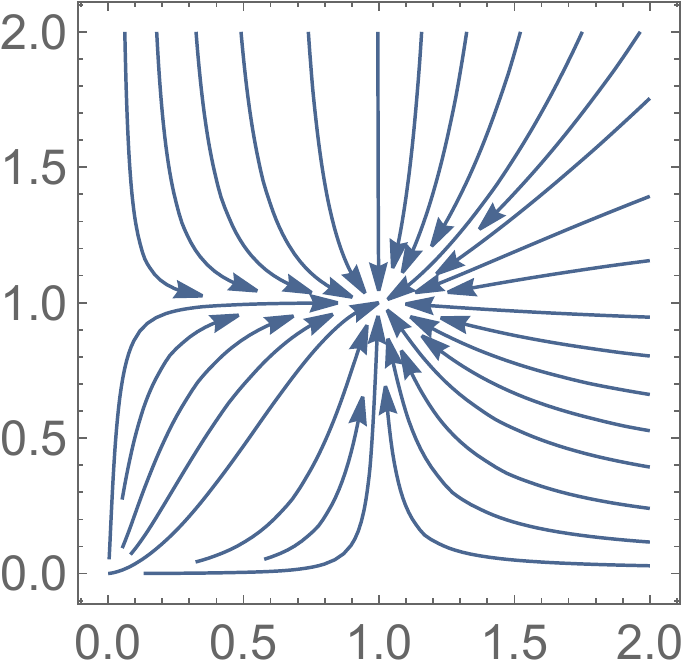}
         \caption{$\mu_1$: ($\gamma=-1/2$) -  $\mu_2$: ($\gamma=-1$).}
     \end{subfigure}
       \hfill
     \begin{subfigure}[b]{0.3\textwidth}
         \centering
         \includegraphics[width=\textwidth]{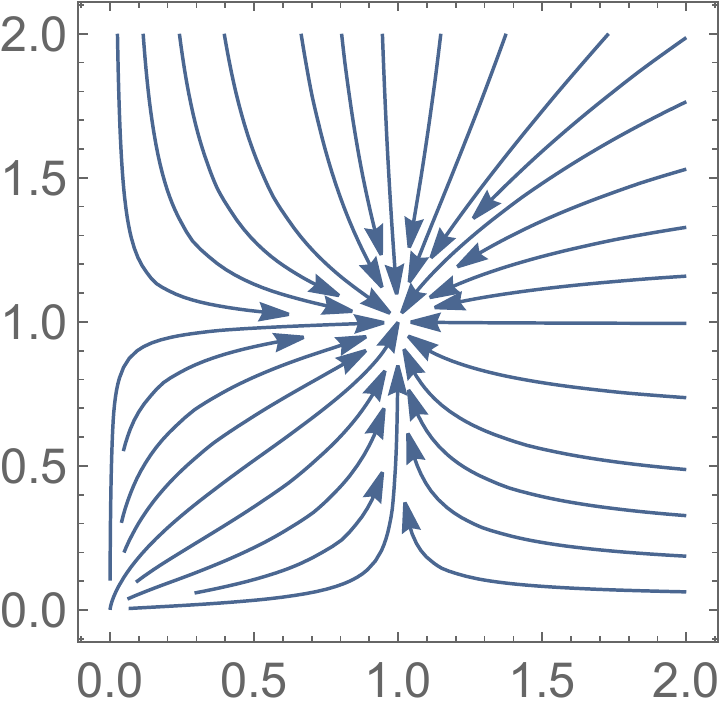}
         \caption{$\mu_1$: ($\gamma=-5/3$) - $\mu_1$: ($\gamma=-1$).}
     \end{subfigure}
          \hfill
     \begin{subfigure}[b]{0.3\textwidth}
         \centering
         \includegraphics[width=\textwidth]{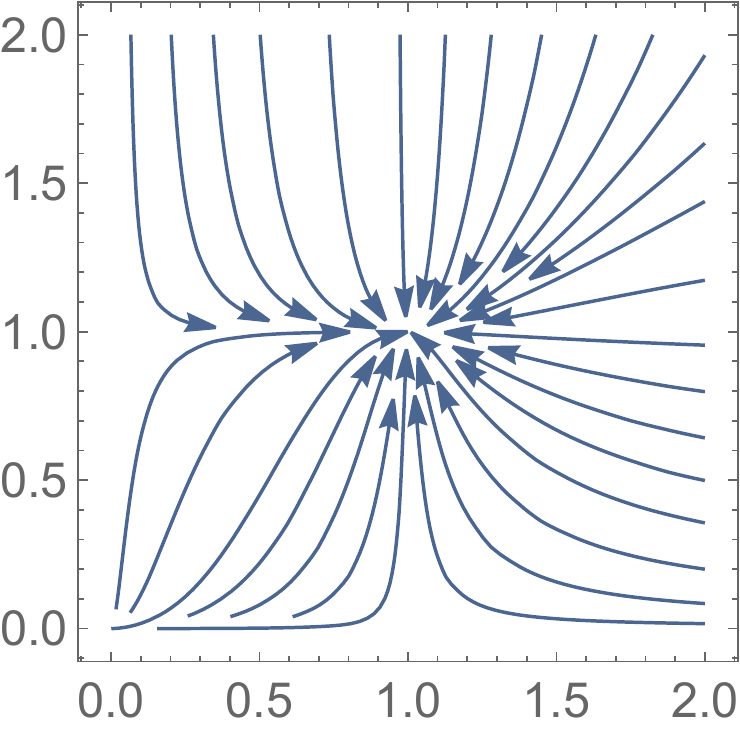}
         \caption{$\mu_1$: ($\gamma=-2/3$) - $\mu_2$: ($\gamma=-5/3$).}
     \end{subfigure}
\caption{($\Omega_1,\Omega_2$)-phase portraits for $\mu_1$, $\mu_2$ in the cases of: (a) ($\gamma=-1/2$)-($\gamma=-1$), (b) ($\gamma=-5/3$)-($\gamma=-1$), and (c) ($\gamma=-2/3$)-($\gamma=-5/3$).}
      \end{figure}

However, using the definition of the reduced vector field and Eq. (\ref{ooc}) that describes the dynamics on the center manifold, we can prove rigorously that the equations (\ref{sys1})-(\ref{sys3}) do not bifurcate.
We know that $f(\Omega,\mu)$ given by (\ref{ooc}) has a dispersive fixed point at the origin because,
\be\label{sn1}
f(0,0)=0,\quad\frac{\partial f}{\partial\Omega}(0,0)=0.
\ee
However, since $f(\Omega,\mu)$ violates the transversality condition, namely, it satisfies,
\be
\frac{\partial f}{\partial\mu}(0,0)=0,
\ee
it follows that $f(\Omega,\mu)$  does not satisfy the conditions for a saddle-node bifurcation. Also because it violates the nondegeneracy condition, that is it satisfies,
\be
\frac{\partial^2 f}{\partial\Omega^2}(0,0)= 0,
\ee
it follows that it cannot bifurcate transcritically.  Finally, because it satisfies the condition,
\be
\frac{\partial^3 f}{\partial\Omega^3}(0,0)= 0,
\ee
it follows that $f(\Omega,\mu)$ cannot satisfy the conditions of the pitchfork bifurcation either.

The conclusion from these results is that because of the lack of bifurcations curves, and the consequent violation of the transversality and the nondegeneracy conditions, the Friedmann system (\ref{sys1})-(\ref{sys3}) cannot properly  bifurcate.


\subsection{The versal unfolding}
We now show that there is \emph{a uniquely defined perturbation} of the system (\ref{ooc})-(\ref{m0}) (and so of the original Friedmann equations  (\ref{sys1})-(\ref{sys3})) that satisfies the conditions of the typical saddle-node bifurcation. We reiterate the fact that in this paper we do \emph{not} consider perturbations of the Friedmann \emph{metric}, but instead study the issue of structural instability of the Friedmann \emph{equations}.

Instead of the one-dimensional field $f(\Omega,\mu)=\mu \Omega(\Omega-1)$ considered above, we now introduce the following \emph{perturbation} of it,
\be\label{un1}
\Omega'=\mu \Omega(\Omega-1)+\sigma,
\ee
where $\sigma$ is a new parameter in a Taylor expansion of the field. We call $\sigma$ \emph{the unfolding parameter} of  (\ref{sys1})-(\ref{sys3}). Although it looks as if (\ref{un1}) has two parameters, we can eliminate $\mu$ by defining a new time,
\be\label{T}
T=\mu\tau,
\ee
to combine the two parameters $\mu\neq 0$ and $\sigma$ into one, which we call the \emph{unfolding-fluid parameter},
\be\label{nu}
\nu=\frac{\sigma}{\mu},
\ee
to get
\be\label{un1a}
\frac{d\Omega}{dT}=\Omega(\Omega-1)+\nu.
\ee
This is a perturbation of Eq. (\ref{ooc}) by adding \emph{lower-order terms} about the dispersive fixed point at the origin, namely,  \textbf{EQ-I}. Then for the new field $f(\Omega,\nu)$ given by the right-hand-side of Eq. (\ref{un1a}), the defining conditions of the saddle-node bifurcation are met: the dispersive fixed point condition (\ref{sn1}), the transversality condition,
\be\label{sn2}
\frac{\partial f}{\partial\nu}(0,0)=1\neq 0,
\ee
and the nondegeneracy condition,
\be\label{sn3}
\frac{\partial^2 f}{\partial\Omega^2}(0,0)=2\neq 0.
\ee
These conditions make the addition of $O(3)$ terms in Eq. (\ref{un1}) redundant because  the field is now determined by the $O(2)$ terms or lower. So the only possible remaining perturbing term now allowed is the $\sigma$ term, thus making the new vector field (\ref{un1a}), or (\ref{un1}),  uniquely defined.

Therefore if we could also show that (\ref{un1a}) is also a \emph{generic family}, it would follow that (\ref{un1}) is a versal unfolding.
The proof proceeds in two steps. The first step is to show that (\ref{un1a}) can be written as the versal deformation of the generic quadratic family. The second step, which is completed at  the end of Section 6, is to suitably shift (\ref{ooc}) to show that it matches with the generic quadratic family.

We introduce the new variable,
\be\label{Z}
Z=\frac{1}{2}-\Omega,
\ee
so that  the system (\ref{un1a}) becomes,
\be \label{un3}
\frac{dZ}{dT}=\bar{\mu}-Z^2,\quad\bar{\mu}=\frac{1}{4}-\nu.
\ee
This is the normal form of the saddle-node bifurcation with parameter $\bar{\mu}$, and bifurcation point at $(Z,\bar{\mu})=(0,0)$. Since it includes the effects coming from perturbation terms of all other orders in the Taylor series, and the saddle-node bifurcation is the versal unfolding of the generic quadratic family,  Step one of the proof follows.

\begin{figure}
\begin{center}
\includegraphics[scale=0.7]{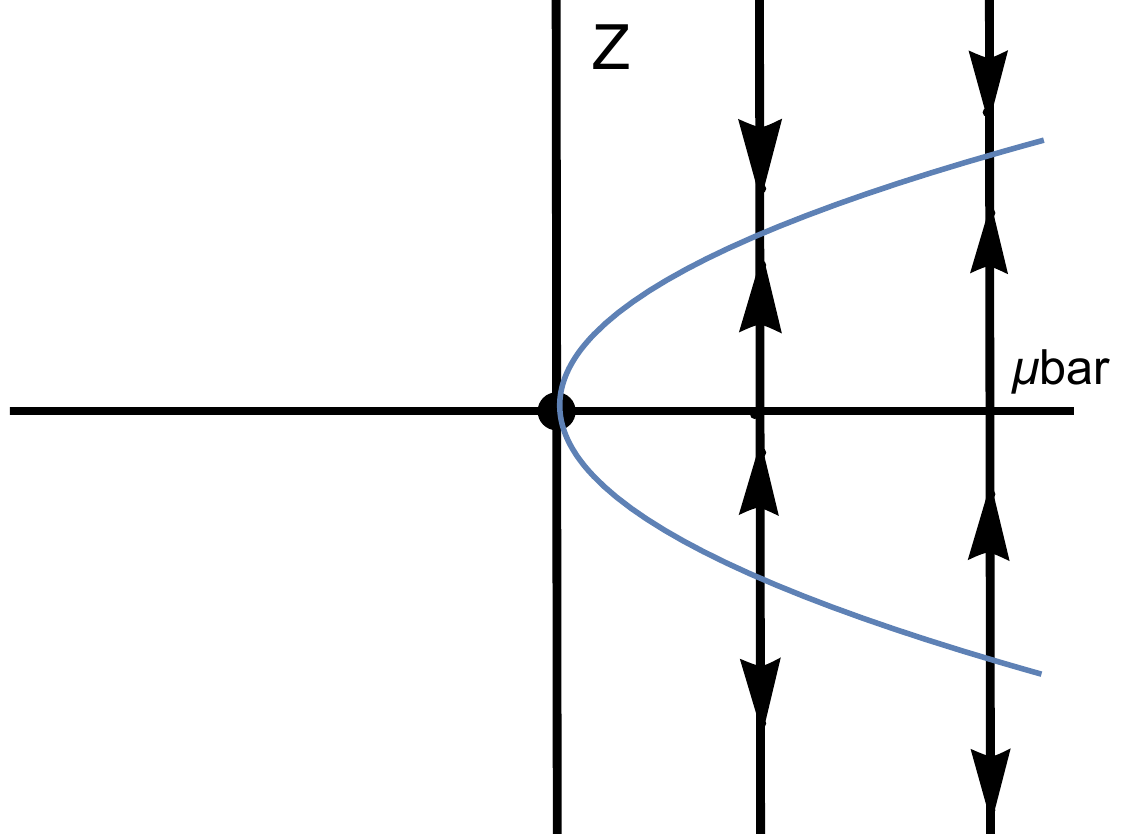}
\end{center}
\caption{Bifurcation  diagram of the system (\ref{un3}) for the evolution of the Friedmannian domain $\mathcal{A}$ (possibly containing causally connected subregions). The stable orbit is in the first quadrant. Phase lines are shown for two different $\bar{\mu}$-values.}
\label{sano1}%
\end{figure}

There is a very great difference in the implications  of the two laws, (\ref{ooc}) and (\ref{un3}), for the structure and evolution of the universe. These become apparent in form of the two bifurcation diagrams in Figs. \ref{b1}, \ref{sano1}, and this has direct implications for the long term behaviour of the cosmological solutions as we now discuss.

In  the case of the standard system (\ref{ooc})-(\ref{m0}), Fig. \ref{b1} shows that for all nonzero values of $\mu$ there are always the two equilibria at $\Omega=0,1$. In particular,  the equilibrium corresponding to $\Omega=1$ implies the presence of an all-encompassing  big bang singularity in the past, that exists for all models.

On the other hand, the bifurcation diagram for the system (\ref{un3}) is the standard one for the normal form of the saddle-node bifurcation, cf. Fig. \ref{sano1}\footnote{It is important at this point that the reader understands the notion of `elementary fixed points', in particular, the  `saddle-node' phase point, cf. e.g., \cite{arny88}, p. 81, and its role in the phase portraits of the saddle-node bifurcation (not shown here).}. When $\bar{\mu}>0$, or $\nu<1/4$, there are two \emph{new} fixed points for the versal unfolding (\ref{un3}) at,
\be\label{zeq}
Z_\pm=\pm\sqrt{\bar{\mu}},
\ee
that is when,
\be \label{un4}
\Omega_\pm=\frac{1}{2}\left(1\pm\sqrt{1-4\nu}\right),\quad\nu<\frac{1}{4},
\ee
(so that in this definition, $\Omega_\pm\gtrless 1/2$), but no fixed points for $\bar{\mu}<0$ (or, $\nu>1/4$).

This means that when the system (\ref{ooc})-(\ref{m0}) unfolds to become (\ref{un3}), the two new equilibria for $\bar{\mu}>0$ - the stable node and the saddle - `merge' at $\bar{\mu}=0$ to become the saddle-node equilibrium (not shown in this figure), only to be annihilated for $\bar{\mu}<0$ (alternatively, the two new fixed point solutions are `created' for $\bar{\mu}>0$ from the saddle-node point at $\bar{\mu}=0$).

Therefore when the system (\ref{un3}) is near the bifurcation curve, all solutions either always approach the stable branch, first quadrant,   or always recede from the negative branch, fourth quadrant, as $T\rightarrow+\infty$. This conclusions are revered in the past time direction.

From Eq. (\ref{zeq}) it follows that the distance between the two equilibria is of the order of $\sqrt{\bar{\mu}}$, and so as $\bar{\mu}\rightarrow 0$ and we approach the  the moment of birth (or `death') of the equilibria (\ref{zeq}),  both of these states approach motion  with an infinite speed (cf. e.g., \cite{arny86}, p. 16).

The case of standard cosmology evolution corresponds to zero unfolding-fluid parameter (or $\sigma=0$), so that $\bar{\mu}=1/4$. In this case, from Eq. (\ref{un4}) we have the standard fixed point solutions  $\Omega=0,1$. For all Friedmann  models,  the flat equilibrium  $\Omega_+$ corresponds to the past (big bang) singularity, cf. \cite{we}, p. 59.

Using the unfolding model (\ref{un3}), however, we can say more: In this picture, both the big bang and the big crunch (when it exists) are created at the moment (in $\tau$-time) when $\bar{\mu}=1/4$, and are at a distance of the order of $1/2$ (in $Z$-units). However,  these equilibria are absent when $\bar{\mu}\neq 1/4$. All other equilibrium pairs (\ref{zeq}), lie in $\Omega$-values in neighborhoods of the point $\Omega=1/2$ of the form $(\Omega_-,\Omega_+)$, with endpoints given by (\ref{un4}). Consequently they experience no big bang or big crunch singularities as the asymptotic limits of the endpoints of their neighborhoods are approached.

In Fig. \ref{sano1}, each vertical line is the phase space of one particular `universe', for instance the standard Friedmann evolution given by
(\ref{ooc})-(\ref{m0}) in general relativity is on the $\bar{\mu}=1/4$-phase line. When the parameter $\bar{\mu}$ passes through its bifurcation value zero, the parabola is created or disappears (if zero is approached by going the other way). In this sense,  all vertical phase lines suddenly appear together with their `prescribed' distances between their fixed points (one stable, one unstable) depending smoothly on $\bar{\mu}$, or the other way,  they disappear because the equilibria combine with one another.

\section{Dispersive type-IIa states}
\begin{figure}
\begin{center}
\includegraphics[scale=0.5]{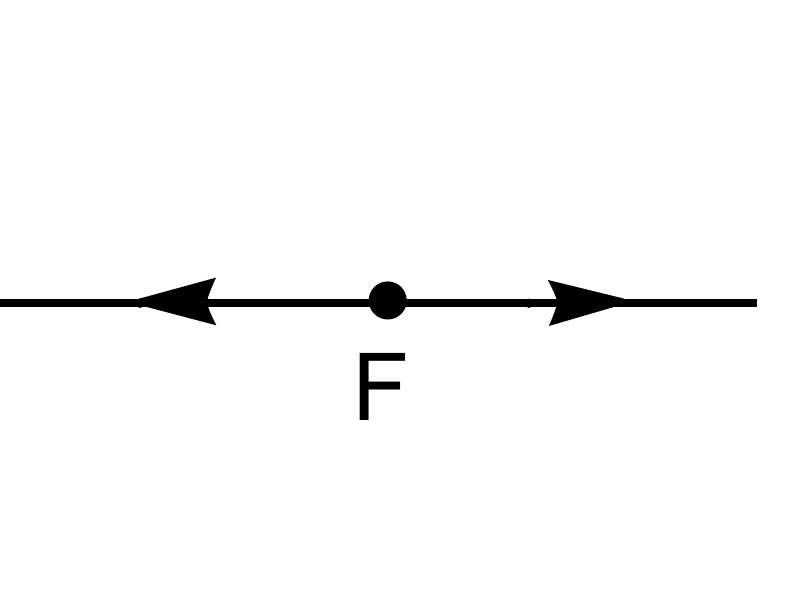}
\end{center}
\caption{The $v$-phase space describing the evolution of all models near the flat equilibrium $F$. }
\label{1d}%
\end{figure}

We now turn to the study of the flat equilibrium solution \textbf{EQ-IIa} of the system (\ref{o0})-(\ref{h0}), and examine the center manifold, and the possibilities of bifurcations and  unfoldings. Because the analysis is completely analogous to that in Section 5, we shall be brief. To write the system in normal form, we first transfer the equilibrium $(1,0)$ to the origin by the linear transformation to new variables,
\be
(v,H):=(\Omega-1,H),
\ee
leading to the new system for the $(v,H)$ variables,
\begin{align}
v'&=\mu v+\mu v^2\label{v0}\\
H'&=-(1+\frac{\mu}{2})H-\frac{1}{2}\mu Hv.\label{v1}
\end{align}
At the origin of the $(v,H,\mu)$ phase-parameter space,  the jacobian of the linearized system is again (\ref{j1}). This leads to the  (\ref{jo1})-form of the equations (\ref{v0})-(\ref{v1}),
\be
\left(
  \begin{array}{c}
    v' \\
    H' \\
  \end{array}
\right)
=
\left(
  \begin{array}{cc}
    0 & 0 \\
    0 & -1 \\
  \end{array}
\right) \left(
  \begin{array}{c}
    v\\
    H \\
  \end{array}
\right)
+
\left(
  \begin{array}{c}
    \mu v+\mu v^2 \\
   -H-\frac{1}{2}\mu H -\frac{1}{2}\mu v H \\
  \end{array}
\right),
\ee
and so to  the form (\ref{jo12a})-(\ref{jo12c}) with $A=0, B=-1, C=0$, namely,
\begin{align}
v'&=\mu v+\mu v^2\label{sys1a}\\
H'&=-H-\frac{1}{2}\mu H -\frac{1}{2}\mu v H \label{sys2a}\\
\mu'&=0.\label{sys3a}
\end{align}
Proceeding as before, we find that  the center manifold is the $v$-line, that is the horizontal line $\Omega=1$, with the evolution equation on it given by the reduced vector field on $W^c_{loc}$,
\begin{align}
v'&=\mu v+\mu v^2\label{cmEQN1}\\
\mu'&=0.\label{m0a}
\end{align}
From the dynamics on the center manifold it then follows that  when $\mu<0$ we have $v'<0$, whereas when $\mu>0$ we have $v'>0$, and we arrive at the phase diagram given in  Fig. \ref{1d}. This shows that the flat state $F$ is a past attractor of all models in the sense that,
\be
\lim_{\tau\rightarrow -\infty} v=1.
\ee
The remaining bifurcation analysis for the vector field (\ref{cmEQN1}) on the center manifold is analogous to that in Sections 5.2 and will not be repeated here. We find that Eq. (\ref{cmEQN1}) does not bifurcate.

However, passing on to the unfolding,
\be\label{un1b}
\frac{dv}{dT}=v+v^2+\nu,
\ee
with the unfolding-fluid parameter $\nu$ given again by (\ref{nu}), and defining
\be \label{W}
W=\frac{1}{2}+v,
\ee
we find the saddle-node bifurcation,
\be \label{un3a}
\frac{dW}{dT}=\bar{\mu}+W^2,\quad\bar{\mu}=\frac{1}{4}-\nu.
\ee
Specifically, the unfolding (\ref{un1b}) is versal. Same comments apply here as those at the end of Section 5.3, in particular, for the motion of the system near   the stable orbit of the parabola,
\be\label{un5}
W_\pm=\pm\sqrt{-\bar{\mu}},
\ee
that is near $W=-\sqrt{-\bar{\mu}}$. We also note that the previous discussion about the long term behaviour and singularities  in the type-I unfolding of the previous Section apply here as well.

We end this Section by completing the proof of the second step that (\ref{un1}) and also (\ref{un1b}) are generic families,  by  reducing  (\ref{ooc}) and (\ref{v0}) to purely   quadratic  vector fields. Both (\ref{ooc}) and (\ref{v0}) are of the form, \be \label{coord}
f(x)=x+ax^2,
\ee
with  $a$  a constant which covers the sign difference in their reductions (\ref{un3}) and (\ref{un3a}) respectively. For any constant $c$, we can perform the coordinate shift, $y=x+c$, which, after substituting in (\ref{coord}) and rearranging the terms in powers of $y$, implies that we can get rid of the term proportional to $y$ if we choose $c=1/(2a)$. This leads to the form
\be
f(y)=-\frac{1}{4a}+ay^2,
\ee
which after another shift becomes,
\be
g(y)=ay^2.
\ee
It is well known that the `deformation' $\dot{y}=a y^2 +\epsilon$ is the versal unfolding of the vector field $\dot{y}=a y^2$,  [A83], p. 167,  [W3], pp. 409-10, from which the result follows.   (The proof of versality relies on showing that it is a generic family: the map $y^2 +\epsilon$  is transversal to the $f(y)$-axis (considered as a submanifold of the plane), and then by an application of  Thom's transversality theorem that `transversal $k$-jet extensions are generic', versality follows).

What we have shown in the  Sections 5, 6 is that the deformations (\ref{un3}),  (\ref{un3a}) unfold versally the Milne and flat states of standard cosmology respectively.

\section{Dispersive type-IIb states}
In this Section we study  the equilibrium solution \textbf{EQ-IIb} of the system (\ref{o0})-(\ref{h0}). There are two different subcases:
\begin{enumerate}
\item \textbf{EQ-IIb-1}: Dispersive Einstein static universe, $(1,0,-2)$: We set $H=0$ in the \textbf{EQ-IIb} line of equilibria $(1,H,-2)$.
\item \textbf{EQ-IIb-2}: Dispersive de Sitter universe, $(1,1,-2)$: We set $H=1$ in the \textbf{EQ-IIb} line of equilibria $(1,H,-2)$. Without loss of generality, this is analogous to the remaining equilibria $H\neq 0$, the treatment is basically the same.
\end{enumerate}
However, the dynamical treatments of the subcases b1, b2 are different and must be done separately.

\subsection{The equilibrium $(1,0,-2)$}
The transformation,
\be
(v,H,\mu_1):=(\Omega-1,H,\mu+2),
\ee
transfers the dispersive Einstein static universe $(1,0,-2)$ to the origin, and leads to  the form (\ref{jo12a})-(\ref{jo12c}) with $A=0, B=-1, C=0$, (after the time-scaling $\tau\rightarrow 2\tau$), namely,
\begin{align}
H'&=-\frac{1}{4}\mu_1 H -\frac{1}{4}\mu_1 H v  \label{sys1b}\\
v'&=-v^2 +\frac{1}{2}\mu_1 v+\frac{1}{2}\mu_1 v^2\label{sys2b}\\
\mu_1'&=0.\label{sys3b}
\end{align}
We note that  here, in distinction to the previous analyses in Sections 5, 6 , the dispersive dimension is $H$.
Proceeding as before, we find that  the reduced dynamics on the center manifold is given by,
\begin{align}
H'&=-\frac{1}{4}\mu_1 H\label{cmEQN2}\\
\mu_1'&=0.\label{m0b}
\end{align}
When $\mu_1>0$ (or, $\mu<-2$), and $H>0$,  we have $H'<0$, and we have stability of the fixed point, whereas when $H<0$ we have $H'>0$, and the solution is unstable. These results describe the stability properties of the dispersive Einstein static universe.

It is also easily checked that due to the violations of the transversality and nondegeneracy conditions, the system (\ref{cmEQN2}) cannot bifurcate.

We shall comment on the question  of  unfolding (\ref{cmEQN2}) at the end of this Section.

\subsection{The equilibrium $(1,1,-2)$}
We now come to  the stability of the dispersive de Sitter space $(1,1,-2)$. The linear change of variables,
\be\label{def2}
(v,w,\mu_1):=(\Omega-1,H-1,\mu+2),
\ee
brings the fixed point $(1,1,-2)$ of the system (\ref{o0})-(\ref{h0}) to the origin, where the system assumes the form,
\begin{align}
w'&=-\frac{\mu_1}{2}+v-\frac{\mu_1 w}{2} -\frac{\mu_1 v}{2}+wv-\frac{1}{2}\mu_1 w v  \label{sys1c}\\
v'&=-2v-2v^2 +\mu_1 v+\mu_1 v^2.\label{sys2c}
\end{align}
For this system, we are interested in the behaviour of solutions near the origin (where also $\mu_1=0$). We note that the dispersive direction (i.e., the $0$-eigenspace) is the $w$-axis, while the $(-2)$-eigenspace is the line $v=-2w$ in the $(w,v)$ phase space.

However, the general conditions on the derivatives that appear in Eq. (\ref{dis}) are incompatible with the presence of the term $-\mu_1/2$ in Eq. (\ref{sys1c}) which is  \emph{linear in the parameter $\mu_1$}. This would then be incompatible with the possible existence of a saddle-node bifurcation at its bifurcation point. So the theory discussed in  Section \ref{par-method} is not applicable in the present case, and we have to proceed differently in order to check for the existence of a saddle-node bifurcation at $\mu_1=0$.

We write the system (\ref{sys1c})-(\ref{sys2c}) in the form,
\be\label{med1}
\left(
  \begin{array}{c}
    w' \\
    v' \\
  \end{array}
\right)
=
\left(
  \begin{array}{cc}
    0 & 1 \\
    0 & -2 \\
  \end{array}
\right) \left(
  \begin{array}{c}
    w\\
    v \\
  \end{array}
\right)
+
\left(
  \begin{array}{c}
    -\frac{1}{2} \\
    0 \\
  \end{array}
\right) \mu_1
+
\left(
  \begin{array}{c}
    G_1 \\
    G_2 \\
  \end{array}
\right),
\ee
where,
\beq
G_1&=& -\frac{\mu_1 w}{2} -\frac{\mu_1 v}{2}+wv-\frac{1}{2}\mu_1 w v,\\
G_2&=&-2v^2 +\mu_1 v+\mu_1 v^2 .
\eeq
Introducing the notation, $z=(w,v)^\top$, and $F(z,\mu_1)=(F_1,F_2)$, for the right-hand-sides of the Eqns. (\ref{sys1c}), (\ref{sys2c}) respectively, the system (\ref{sys1c}), (\ref{sys2c}) is written as $z'=F(z,\mu_1)$, and so  (\ref{med1}) is of the general  form,
\be
z'=D_z F(0,0)z+D_{\mu_1} F(0,0)\mu_1 +G(z,\mu_1),
\ee
where $G=(G_1,G_2)^\top$.
What we have gained presently is that the term  $D_{\mu_1} F(0,0)\mu_1$ that appears here, was zero previously, that is under the assumptions (\ref{dis}) for the system (\ref{sys1c}), (\ref{sys2c}).

In this new formulation,  we can bring the system (\ref{med1}) to having its linear part in canonical form as follows.
The eigenvector matrix for the linearized Jacobian of (\ref{sys1c})-(\ref{sys2c}) is,
\be
T=\left(
  \begin{array}{cc}
    1 & -1 \\
    0 & 2 \\
  \end{array}
\right),
\ee
and so introducing the new variables,
\be \label{trans1}
z=T\left(
  \begin{array}{c}
    x \\
    y \\
  \end{array}
\right),
\ee
we find that the system (\ref{med1}) takes the form,
\be\label{med2}
\left(
  \begin{array}{c}
    x' \\
    \mu_1' \\
    y'\\
  \end{array}
\right)
=
\left(
  \begin{array}{ccc}
    0 & -1/2 & 0 \\
    0 & 0 & 0 \\
    0 & 0 & -2\\
  \end{array}
\right) \left(
  \begin{array}{c}
    x\\
    \mu_1 \\
    y\\
  \end{array}
\right)
+
\left(
  \begin{array}{c}
   f \\
    0 \\
    g\\
  \end{array}
\right),
\ee
where,
\beq
f&=& -\frac{\mu_1 w}{2} +wv-\frac{1}{2}\mu_1 w v-v^2+\frac{\mu_1 v^2}{2},\\
g&=&-v^2 +\mu_1 v+\mu_1 v^2 .
\eeq
Eq. (\ref{med2}) is a block diagonal system for its linear part, with the zero eigenvalues coming first and the non zero eigenvalue after the zero ones. Therefore we can now proceed with the calculation of the parametric center manifold and set,
\be
y=h(x,\mu_1)=ax^2+bx\mu_1+c\mu_1^2 +O(3).
\ee
After some algebra, we find that the dynamics on the center manifold is described by,
\be
x'=-\frac{1}{2}\mu_1-\frac{1}{2}\mu_1 x,
\ee
or, using the definition (\ref{trans1}), and the fact that on the center manifold, $v=0$, we find
\be
w'=-\frac{1}{2}\mu_1-\frac{1}{2}\mu_1 w,
\ee
or using (\ref{def2}),
\be \label{cmEQN3}
H'=-\frac{1}{2}\mu_1 H.
\ee
This equation is to be compared with Eq. (\ref{cmEQN2}), the corresponding one for the other type II-b state, they are the same, and so similar stability conclusions apply presently.

We end this Section with a comment about the two families (\ref{cmEQN2}), (\ref{cmEQN3}). The vector field $f(\mu_1,H)=\mu_1 H$ is an unfolding of the zero vector field $f(0,H)=0$, for any $H$. However,  it is not a \emph{versal} unfolding of $f(0,H)$, for instance the unfolding $g(\mu_1,H)=\mu_1$ does not have a fixed point for $\mu_1\neq 0$, and therefore is not conjugate to $f(\mu_1,H)$ which has an equilibrium at zero. This argument shows that unlike the type I and IIa cases, the  type II-b solutions are rather special because they cannot be versally deformed to give generic  unfoldings.

\section{Applications to synchronization}
We now return to the problem stated in the Introduction, namely,
\begin{enumerate}
\item \emph{Question 1}: Does a single Friedmann domain that is initially synced and homogenized remain so during its evolution?
\item \emph{Question 2}: Do two causally disconnected Friedmann domains   sync during their evolution if they were not so synced initially?
\end{enumerate}
For both of these questions, we shall assume that the standard theorems on local existence and uniqueness, extension to a compact set containing the initial condition, and differentiability of solutions with respect to parameters (cf. e.g., [W3], pp. 90-91) hold for each one of the systems we consider for sync.

In Section 8.1, we derive the evolution law of a set of causally disjoint subdomains after manifold reduction. A description of dynamical synchronization in presented in  Section 8.2 (for more details and generality, the reader may consult  \cite{sync1} and refs. therein). In Sections 8.3-8.6, we examine for synchronization evolving Friedmann domains corresponding to dispersive solutions and their unfoldings.

\subsection{Common evolution of a pair of uncorrelated subdomains}
We imagine that the Friedmannian domain $\mathcal{A}$ is the union of an arbitrary number of Friedmannian subdomains $\mathcal{A}_i$, that is  $\mathcal{A}=\cup_{i\in I}\mathcal{A}_i$. We shall now show that we can  take without loss of generality the single common set of  equations (\ref{ooc})-(\ref{m0}) to be the set of equations that govern  the evolving Friedmann distribution of causally disjoint domains after center manifold reduction. That is,  the distribution of the evolving causally disjoint domains satisfies  the common system  (\ref{sys1})-(\ref{sys3}), with phase space  the space of states $(\Omega,H,\mu)$, where  different phase points  denote different subdomains $\mathcal{A}_i$.

Let us first  suppose that we have two causally disjoint  spatial domains $\mathcal{A}_1$, $\mathcal{A}_2$  that evolve according to \emph{different} dynamical laws  given by:

\noindent\textbf{Domain} $\mathcal{A}_1$ :
\begin{align}
\frac{dH_1}{d\tau_1}&=-(1+q_1)H_1\label{h1} \\
\frac{d\Omega_1}{d\tau_1}&=-\mu_1 (1-\Omega_1)\Omega_1
\label{o1}\\
q_1& =\frac{1}{2}\mu_1\Omega_1,\label{q1}
\end{align}
\textbf{Domain}  $\mathcal{A}_2$:
\begin{align}
\frac{dH_2}{d\tau_2}&=-(1+q_2)H_2\label{h2} \\
\frac{d\Omega_2}{d\tau_2}&=-\mu_2 (1-\Omega_2)\Omega_2
\label{o2}\\
q_2& =\frac{1}{2}\mu_2\Omega_2,\label{q2}
\end{align}
where we have set $\mu_i=3\gamma_i-2,i=1,2$. Here the two causally disjoint  fluid-filled domains have Hubble parameters $H_i$, density parameters $\Omega_i=\rho_i/(3H_i)$, equations of state  $p_i=w_i\,\rho_i, i=1,2$,  $w_i=\gamma_i-1$,  and  we use the dimensionless times $\tau_i$ instead of the proper time $t$, given by,
\begin{equation}
\label{dimless2}
\frac{dt}{d\tau_i}=\frac{1}{H_i},
\end{equation}
defined  through the scale factors by
\begin{equation}
\label{dimless1}
a_i=a_{i,0}\, e^{\tau_i}.
\end{equation}
We consider only the case $\gamma_1\neq \gamma_2$ in the two domains, because $\gamma_1=\gamma_2$  implies that,
\be
H_1=H_2\Leftrightarrow\Omega_1=\Omega_2,
\ee
and we are back to the single domain problem.

To examine the two domains for sync (see also below for the general case of an arbitrary number  of domains), we assume that the spatial domains $\mathcal{A}_1$, $\mathcal{A}_2$ \emph{transmit} through light signals part of the information of their states to the common future point $G$, and  take the transmitting parts of the signals from $\mathcal{A}_1$ and $\mathcal{A}_2$ to $G$ to be $H_1, H_2$ respectively.

This choice is in line with our previous results on center manifold reduction of the Milne and flat dispersive states, where we basically showed that one may ignore the $H$-equation for the local dynamics, and restrict attention to the reduced dynamics on the center manifold, the system is (\ref{ooc}), (\ref{m0}).

We then simply  set $H_2=H_1$ in Eq. (\ref{h2}), and examine the behaviour of the remaining part of the solution $(\Omega,H)$, namely, the $\Omega$-part. Dividing Eqns. (\ref{o1}), (\ref{o2}), and integrating,  we find that either $\Omega_1=\Omega_2$, or, $\Omega_1=2-\Omega_2$. It then follows by a simple calculation that  if Eq. (\ref{o2}) holds in the domain $\mathcal{A}_2$, then Eq. (\ref{o1}) must also hold  on  the \emph{same} domain $\mathcal{A}_2$, and vice-versa. Hence, the common system of equations  (\ref{sys1})-(\ref{sys3}) holds throughout both of the  Friedmann domains $\mathcal{A}_1$, $\mathcal{A}_2$ with the understanding that in the two subdomains  the density parameters $\Omega_i,i=1,2$ may be translations of the function $\Omega=\Omega_1$.

\subsection{General synchronization}
The problem of Question 2 for a distribution of an arbitrary number of  causally disconnected domains $\mathcal{A}_i,i\in I,$ is likewise reduced to that of examining whether a single domain $\mathcal{A}$ governed by the  equations (\ref{sys1})-(\ref{sys3}), in line with center manifold reduction.

Let us consider an \emph{arbitrary} number of  transmitters, $x_i=(\Omega_i,H_i,\mu_i),i\in I$,  the causally disjoint distribution of subdomains $\mathcal{A}_i,i\in I$ inside the single Friedmann domain $\mathcal{A}$, each sending light signals  eventually arriving at the common `receiver' sitting at the spacetime point $G$ in their future overlooking them (this is analogous to the situation arising in the horizon problem).

We start with the system (\ref{o0})-(\ref{h0}), and the vector variable $x=(\Omega,H)\in \mathbb{R}^2$, and consider  the `transmitter':
\be\label{a1}
\dot{x}=f(x),
\ee
with the $\cdot\equiv d/d\tau$, and  $f(x)=(-\mu\Omega+\mu\Omega^2,-H-\mu H\Omega/2)$.

To examine the distribution of domains  $\mathcal{A}_i,i\in I$ for sync, we proceed as in the next four steps:

\noindent\textsc{Step 1:} Split the transmitter in two subsystems $a, b,$ with the indexing set written as $I=a\cup b$,  by setting  $x_i=(u_a,v_b)$, with $u_a=\Omega_a,v_b=H_b$, the $a$-variables are denoted by $u$, and the $b$-variables by $v$. Then  (\ref{a1}) becomes,
  \bq
  \dot{u}_a&=&f_a (u_i,v_i),\label{sysA} \\
  \dot{v}_b&=&f_b (u_i,v_i),\label{sysB}.
  \eq
Here we are assuming a common time for all subsystems.

\noindent\textsc{Step 2:} The receiver at $G$ receives only the `$v$-part' of the signals from the transmitter, and  at $G$ we set,
\be
v_b=v,\, \quad(\textrm{or,}\,\, H_b=H).
\ee

\noindent\textsc{Step 3:} Using the remaining equations, the `$u$-part' of the system (\ref{sysA}), (\ref{sysB}), namely,
$\dot{u}_a=f_a (u,v),
$ the receiver then  checks the difference in the \emph{flows} of the remaining $a$-variables, namely, the \emph{sync function},
\be
\omega=|u_{a_i}-u_{a_j}|,\, a_i,a_j\in a,\, i\neq j.
\ee

\noindent\textsc{Step 4:} The single domain  $\mathcal{A}$ is said to \emph{sync in the future} (resp. past) provided the orbits of the system (\ref{sysA}) satisfy,
  \be
  \omega\rightarrow 0, \,\,as\,\, \tau\rightarrow\infty\, (\textrm{resp.} -\infty).
  \ee
In this case,   $\mathcal{A}$    evolves keeping its causally disjoint subdomains  in perfect unison.

In the presence of parameters, the above procedure extends  analogously to parameter-dependent flows.

The almost magical property of dynamical systems   that have the ability to sync with each other is that although the receiver has received only part of the information of the transmitters (the `$v$-part'),  it somehow manages to reconstruct the remaining `$u$-part'.

\subsection{Friedmann synchronization}
As we already discussed earlier, the above procedure complies perfectly with our center manifold reduction, through which we arrived at the reduced equation (\ref{ooc}) holding on the center manifold. To introduce a suitable sync function for our problem, we therefore proceed as follows.

Let us start with the general system (\ref{gena})-(\ref{genb}) and set $x=\Omega$,  and $\phi_\tau (\Omega,\mu)$ for the flow of the reduced system (this covers subsequent cases like the system (\ref{ooc})-(\ref{m0}), or the various unfoldings on the center manifold). We  define the \emph{sync function} $\omega(\Omega,\mu)$ to be,
\be\label{sync-def0}
\omega(\Omega,\mu)=|\phi_\tau (\Omega_+,\mu_+)-\phi_\tau (\Omega_-,\mu_-)|,
\ee
where $\Omega_\pm,\mu_\pm$ denote any pair of states of $\Omega$ and any two values of $\mu$ respectively. Usually, in bifurcation problems, $\mu_\pm$ denote values of the bifurcation parameter on either side of its bifurcation value, and similarly for the states $\Omega_\pm$.
The sync function checks for traces of any overall difference in the density function solutions as  functions of $\mu$ during the evolution of the domain, usually as $\mu$ passes through its bifurcation value $\mu=0$.

We now have the following definition.
\begin{definition}[Sync for Friedmann domains]We say that the domain   $\mathcal{A}$ is \emph{dynamically synchronized}, or \emph{remains synced during future evolution}, provided that for all $\mu\in\mathbb{R},\Omega>0$, we have:
\be\label{sync-def1}
\lim_{\tau\rightarrow\infty}\omega(\Omega,\mu)= 0.
\ee
\end{definition}

According to this definition, the domain $\mathcal{A}$ \emph{desynchronizes} during evolution (in the past or future) either when the limit in (\ref{sync-def1}) does not exist, or,  if there are two values $\mu_\pm$ and/or phase points $\Omega_\pm$ for which the sync condition  (\ref{sync-def1}) is violated. This definition is stated in future terms, but sync in the past evolution simply corresponds to instead taking the $\tau\rightarrow-\infty$ limit in the above definitions.

\subsection{Sync and proper-time sync}
Before we proceed further with the sync problem, it is interesting to compare the sync definition for subdomains introduced here  to the more common notion of proper-time synchronization in general relativity.
We shall show now that the two are different, and that two proper time synchronizable subdomains governed by the Eqns. (\ref{o1}), (\ref{o2}), cannot be  \emph{dynamically} synced to each other during their evolution.

Let us assume that the two domains $\mathcal{A}_1$, $\mathcal{A}_2$ are \emph{proper-time synchronizable}, that is there is a common proper time function $t$ for both. Using Eq.  (\ref{dimless2}), we can introduce the common dimensionless time $\tau=\tau_1=\tau_2$, and from  Eqns.  (\ref{h1}), (\ref{h2}), we then get $q_1=q_2$.

This assumption  i.e., common proper time for all uncorrelated domains - subdomains of $\mathcal{A}$),  is a rather strong one because in this way  any observer field $U$ on $\mathcal{A}$  (that is any timelike, future-pointing unit vector field,  parametrized by proper time $t$), becomes geodesic and irrotational, and such that $U=-\textrm{grad}\, t$, \emph{across all subdomains} of $\mathcal{A}$. This assumption thus implies a fine-tuning of all observer fields across $\mathcal{A}$, and so it is rather special.

However, this problem can be easily rectified by defining $t$ to be not the proper time but a kind of average time over all observers in $\mathcal{A}$, so that there exists a smooth function $p>0$ on $\mathcal{A}$ with $U=-p\,\textrm{grad}\, t$. In this way, the elapsed proper time between two spacelike hypersurfaces normal to $U$ will be different in different $\mathcal{A}_i$'s (that is different $U$ observers). In this case, $U$ is just called `synchronizable', rather than proper time synchronizable (cf.  \cite{sw}, p. 54 for this terminology where our `average time' is called a `compromise time').

Dynamical sync as defined earlier requires the validity of the further condition (\ref{sync-def1}) that appears  in the sync definition above, and so is more subtle than  the standard  proper time sync. The former refers to a property of the flow of the Einstein equations (\ref{o0}), (\ref{h0}),  whereas the latter to the flow of the $\partial_t$ vector field being geodesic and irrotational.

In the remaining of this Section, we shall examine how dynamical sync  applies to the dispersive states of Type I, II.

\subsection{Do dispersive type-I states sync?}
Looking back at the diagram in Fig. \ref{b1} having the above definition of Friedmann domains  in mind, it becomes almost obvious to expect that the sync condition  \ref{sync-def1} would be violated in this case, and so  the model (\ref{ooc})-(\ref{m0}) is expected to desync\footnote{It is interesting to note that he same expectation obviously applies to the dynamics near the hyperbolic equilibria
\textbf{EQ1}, \textbf{EQ2}.}. We can show this rigorously as follows.

From (\ref{ooc}), it follows that  for any $\mu=\mu_-<0$, the dynamics will be qualitatively the same as that of the system,
\be\label{x1}
x'=x-x^2,
\ee
and for any $\mu=\mu_+>0$ the dynamics of (\ref{ooc}) will be  equivalent to that of the vector field,
\be\label{x2}
x'=x^2-x.
\ee
(We note however, that the two vector fields in each  pair will not be $C^k$-\emph{conjugate}, $k\geq 1$, to each other except when $\mu_\pm =\pm 1$.) Here, we used the same symbol $x$ to denote $\Omega$ in these ranges of $\mu$ (cf. the diagram in Fig. \ref{b1}).

If we denote the flows of (\ref{x1}) and (\ref{x2}) by $\phi_\tau,\psi_\tau$ respectively, then an easy calculation gives,
\be
\phi_\tau (x)=\frac{xe^\tau}{xe^\tau-x+1}
,\quad
\psi_\tau (x)=\frac{x}{x-xe^\tau+e^\tau},
\ee
with $\phi_\tau (0)=0, \phi_\tau (1)=1$, and $\psi_\tau (0)=0, \psi_\tau (1)=1$, at the fixed points. Then the corresponding velocity fields of the flows are given by,
\be
\frac{d\phi_\tau}{d\tau}= \frac{xe^\tau(1-x)}{(xe^\tau-x+1)^2},
\quad \frac{d\psi_\tau}{d\tau}=\frac{xe^\tau(x-1)}{(x-xe^\tau+e^\tau)^2}.
\ee
For those states that satisfy $x<1$, the velocity fields satisfy,
\be
\frac{d\phi_\tau}{d\tau}>0, \quad\textrm{and}\quad \quad \frac{d\psi_\tau}{d\tau}<0,
\ee
or,
\be
\frac{d\phi_\tau}{d\tau}-\frac{d\psi_\tau}{d\tau}>0,
\ee
that is the function,
\be
\phi_\tau (x)-\psi_\tau (x),
\ee
is increasing for $x<1$.

This result  implies that there are states for which the  sync function $\omega(\mu_-,\mu_+)$ cannot satisfy the sync definition (\ref{sync-def1}), and therefore the Friedmann domain $\mathcal{A}$ desynchronizes during its future evolution in this case.

\subsection{Sync for unfoldings of the dispersive type-I states}
Let us now examine the versal unfolding (\ref{un3}) for sync. Since the system bifurcates at $\bar{\mu}=0$, we need to examine each of the two branches of the bifurcation curve in Fig. 6 separately.

In the unstable branch, the flows on each phase line for each $\mu$ have the property of moving all nearby phase points away from the branch, and so the difference of any two such flows may not tend to a well-defined limit during  evolution. Therefore the sync function will not tend to a definite limit, and so the system will desync in this case.

However, let us consider the stable orbit of the bifurcation curve in Fig. 6. Introducing the variable,
\be
w=\frac{Z}{\sqrt{\bar{\mu}}},
\ee
we find after some calculation that the flow $\zeta_T (w)$ of  Eq. (\ref{un3}) is given by,
\be
\zeta_T (w)=\frac{(1+w)e^T -(1-w)}{(1+w)e^T +(1-w)},
\ee
so that the velocity of the flow is,
\be
\frac{\partial\zeta_Y}{\partial T}=\frac{2(1-w^2)e^T}{((1+w)e^T +(1-w))^2}.
\ee
This implies that $\zeta_T$ is increasing (decreasing) on $|w|<1 (w>1)$, a result to be intuitively expected because if $\bar{\mu}\lessgtr Z^2$, then  $dZ/dT\lessgtr0$, as it follows from the bifurcation diagram in Fig. 6.

We therefore find that $\omega(\bar{\mu}_1,\bar{\mu}_2)$ satisfies the definition (\ref{sync-def1}),
\be
|\zeta_T (w_1)-\zeta_T (w_2)|\rightarrow 0, \quad T\rightarrow+\infty,
\ee
for any $\bar{\mu}_1, \bar{\mu}_2>0$, and for any pair of states $w_1,w_2$ on the same phase line and near any point on the stable branch on either side of the fixed point $w=1$.

Therefore the Friedmann domain $\mathcal{A}$ \emph{governed by the unfolding} (\ref{un3}), or, what is equivalent, (\ref{un1a}), synchronizes as $T\rightarrow+\infty$. From Eq. (\ref{T}), this direction is attained when the $\textrm{sign}(\mu\tau)>0$, and it can happen either in the past or in the future in $\tau$-time.  Since the system bifurcates at $\bar{\mu}=0$,  it will either follow the stable or the unstable branch depending on whether the system finds itself near the one or the other of its two fixed points.

\subsection{Sync and  dispersive type-IIa states}
Using a similar approach like in the previous subsection, it is evident how to proceed in order to examine  the reduced vector field (\ref{cmEQN1}) as well as for its unfolding (\ref{un3a}) for sync, and so we shall be brief. The results are similar to those achieved for the type-I states sync problem.

The original vector field (\ref{cmEQN1}) does not sync and the proof is completely analogous to that given in Section 8.3, but now we are interested in the behaviour of states near to the $\Omega=1$ line, instead of those on the $\Omega=0$ line. For $\mu>0$, the vector field is qualitatively equivalent to $\dot{x}=x+x^2$, with flow given by the form,
\be
\phi_\tau (x)=\frac{x}{(1+x)e^{-\tau}-x},
\ee
while for $\mu<0$, we find
\be
\psi_\tau (x)=\frac{x}{(1+x)e^{\tau}-x}.
\ee
These results imply that the function $\phi_\tau-\psi_\tau$ is increasing for states with $x>0$. We conclude that the limit condition (\ref{sync-def1}) is violated and therefore Friedmann domains cannot sync during evolution in this case.

It is not difficult to see using the method of the previous subsection that for the unfolding (\ref{un3a}), sync is achieved during evolution near the stable branch of the orbit which is now located at the third quadrant of the $(W,\bar{\mu})$-plane.

\subsection{Sync and dispersive type-IIb states}
For reasons completely analogous to those discussed in previous subsections of this Section, we conclude that in the case of  the type-IIb solutions the model cannot sync. Also for reasons described in Section 7, they cannot unfold to describe synced states. Therefore we may regard type II-b solutions as non-generic manifestations that cannot unfold the degenerate solution $H=0$. The main reason is the absence of quadratic terms of the form $H^2$ in the center manifold reduced vector fields.

\section{Discussion}
In this paper we introduced and  studied  two new factors that affect  the dynamics of cosmological models. These are the consideration of \emph{parameters} - as opposed to constants -  in the basic equations that govern the evolution of the universe, and the appearance of \emph{zero eigenvalues} in the linear part of the vector field that defines these equations. These factors  challenge basic aspects of standard cosmology,  and prove that restricting attention to only hyperbolic fixed points is inadequate for the complete treatment of cosmology.

Consequently, we are lead by necessity to the wider consideration of  dispersive  methods in cosmological dynamics, where the newer fields of bifurcation theory, singularity theory, and their applications come into play and need to be taken seriously. At a most basic level, this view requires the consideration of not the usual `individual' system of Friedmann equations, but a whole \emph{family} of such systems parametrized by the fluid parameter.

We have shown how dispersive methods can be used to study  a Friedmann universe with a perfect fluid, where the fluid parameter plays the role of a continuous parameter that may vary. Our results provide the first example of the new possibilities that may arise in this respect. We have shown how the phase portraits as well as the Friedmann equations \emph{themselves} qualitatively change when the fluid parameter varies continuously and smoothly.
These changes are not random or artificially inflicted by hand, but are governed and dramatically constrained by the form of the linear part of the vector field that defines the equations.

The first question to be addressed in this program is whether the solutions of the Friedmann equations  have any bifurcation properties. We have shown that although we are led to the presence of new dispersive (i.e., non-hyperbolic) equilibria, namely, the dispersive versions of the Milne, flat, Einstein static, and de Sitter spacetimes, the field cannot properly  bifurcate. This is due to the absence of a genuine bifurcation \emph{set} in the Friedmann equations, as can be most clearly seen in  the `bifurcation diagram' of the Friedmann cosmological equations.

A further necessary technical advance is the computation of the resulting center manifolds corresponding to the new dispersive equilibria. We calculated center manifolds for all such fixed points, and showed how the overall dynamics of the systems  is usefully reduced on them. From the stability properties of the solutions that lie nearby the dispersive equilibria and on the center manifolds, the (lack of) bifurcation properties of the associated systems could then be readily deduced.

The next question was whether there is some unfolding of the vector fields having the dispersive equilibria, and if yes, whether those could properly bifurcate. We showed the existence of a versal unfolding for the dispersive Milne and flat cases, and computed its properties. We proved that these two vectors fields unfold versally like in the normal form of the  saddle-node bifurcation. Consequently, we predicted the existence of new cosmological solutions for the positive range of the bifurcation parameter for the Milne and flat solutions. On the other hand, the Einstein static universe as well as the de Sitter equilibria cannot unfold this particular system because they are not sufficiently degenerate.

These results have  interesting applications for the long term behaviour of the solutions and the question of singularities, as well as  in the horizon problem. Contrary to the situation in standard Friedmann fluid cosmology where there is a big  bang singularity in the past for all models, the case of the versal unfoldings reveals novel and intriguing possibilities for the past and future evolution of the universe. These were discussed in Sections 5, 6.

The  question of synchronization  required for a resolution of the horizon problem was discussed in Section 8. We showed that there is a synchronization function which asymptotically vanishes in the future for the versal unfoldings  of the Milne and the flat equilibria. Therefore in these two cases, Friedmann universes  completely synchronize in the future. This constitutes a new approach to the horizon problem independently of any consideration of an initial inflationary phase.

There are various directions towards which this work may be usefully extended. Throughout this work we have neglected the possible effects  of a cosmological constant as it leads to a drastic change in the complexity of the problem because of the presence of a \emph{second} parameter (besides the fluid parameter) in the dynamical equations (the $\Lambda$). Keeping the fluid parameter fixed, a cosmological constant would dramatically  change the structure of the Jordan canonical form of the linear part of the vector field, thus probably extending the present results to the next level of degeneracy. Of course other cosmological problems would similarly lead to more degenerate dynamical systems by increasing the codimension of the bifurcation.

An further interesting question related to the present results is about the nature of the underlined gravity theory that contains a perfect fluid as matter source and leads directly to the unfolding (\ref{un1}). An answer to this question is required in order to understand the physical meaning of the unfolding parameter $\sigma$, or its dimensionless version $\nu$.

\section*{Acknowledgments}
The author is  grateful to three anonymous referees for many instructive comments that led to an improvement of this work. The author thanks Panagiota Kanti, Nick Mavromatos, Jose Mimoso, and John Miritzis for discussions on an earlier version of this paper, and he is a especially grateful to Gary Gibbons for a discussion on the unfolding parameter, and for his precious time devoted to  reading the final manuscript. A visit of the author to the DPMMS, University of Cambridge, was made possible after the kind invitation of Mihalis Dafermos, where many helpful discussions with him and also with John Webb took place, all of these are gratefully acknowledged.   This research  was funded by RUDN university,  scientific project number FSSF-2023-0003.


\section*{Declaration}
The authors has no relevant financial or non-financial interests to disclose.

\section*{Data availability statement}
The datasets generated during and/or analysed during the current study are available from the corresponding author on reasonable request.


\begin{thebibliography}{99}
\bibitem{cy22}S. Cotsakis and A. P. Yefremov, Phil. Trans. R. Soc. A380 (2022) 20210191 (Part A); \emph{ibid.}, 20210171, (Part B); arXiv:2203.16443
\bibitem{weinberg2}S. W. Weinberg, \emph{Cosmology} (OUP, Oxford, 2007)
\bibitem{we}J. Wainwright and G. F. R. Ellis, \emph{Dynamical systems in cosmology} (CUP, Cambridge, UK, 1997)

\bibitem{hs}M. W. Hirsch and S. Smale, \emph{Differential equations, dynamical systems and linear algebra} (Academic Press, 1974)
\bibitem{arny94}V. I. Arnold, V. S. Afrajmovich, Yu. S. Il'yashenko, and L. P. Shilnikov, \emph{Bifurcation theory and catastrophe theory}, In: \emph{Dynamical Systems V}, V. I. Arnold (Ed.) (Springer, 1994)

\bibitem{wie58}Norbert Wiener, \emph{Cybernetics} (MIT Press, Cambridge, Massachusetts, 1961)
\bibitem{win67}A. T. Winfree, The geometry of biological time (Springer-Verlag, New York, 1980)
\bibitem{pes75}C. S. Peskin, Mathematical aspects of heart physiology (Courant Institute Publications, New York, 1975), pp. 258-268
\bibitem{ku75}Y. Kuramoto, in H. Araki (ed.), Lecture Notes in Physics, \emph{International
Symposium on Mathematical Problems in Theoretical Physics}, vol. 39. (Springer-Verlag, New York, (1975)), p. 420
\bibitem{ku84a}Y. Kuramoto, \emph{Chemical Oscillations, Waves, and Turbulence} (Springer-Verlag, New York, (1984))
\bibitem{ku84b}Y. Kuramoto, Progr. Theor. Phys. Suppl. 79 (1984) 223
\bibitem{pc} L. M. Pecora, T. L. Carroll, , Phys. Rev. Lett. 64, 821 (1990)
\bibitem{stro-mir90}R. E. Mirollo and S. Strogatz, SIAM J. App. Math. 50 (1990) 1645
\bibitem{stro03}S. Strogatz, \emph{Sync, the emerging science of spontaneous order} (Penguin, 2003)
\bibitem{elp}D. Eroglua, J. S. W. Lamb, and T. Pereira, Cont. Phys. 58, 207 (2017)
\bibitem{mtw}C. W. Misner, K. S. Thorne, J. A. Wheeler, \emph{Gravitation} (Freeman, New York, 1973).
\bibitem{di-pee}R. H. Dicke and P. J. E. Peebles, \emph{The big-bang cosmology-enigmas and nostrums}, In: General Relativity, An Einstein centenary survey, S. W. Hawking and W. Israel, Eds. (CUP, Cambridge, 1979).
\bibitem{li90}Linde, A. (1990). \emph{Particle physics and inflationary cosmology} (Vol. 5). CRC press.
\bibitem{weinberg1}S. W. Weinberg,  Gravitation and Cosmology (John Wiley and Sons, 1972)
\bibitem{ba20}J. D. Barrow, Phys. Rev. D102 (2020) 024017; arXiv: gr-qc/2006.01562
\bibitem{sync1}S. Cotsakis,  Phil. Trans. R. Soc. A380 (2022) 20210189; arXiv: 2010.00298


\bibitem{arny83}V. I. Arnold, \emph{Geometrical methods in the theory of ordinary differential equations,} 2nd Edition (Springer, 1983)
\bibitem{guck-ho83}J. Guckenheimer and P. Holmes, \emph{Nonlinear oscillations, dynamical systems, and bifurcations of vector fields} (Springer, 1983)
\bibitem{wig}S. Wiggins, \emph{Introduction to applied nonlinear dynamical systems and chaos,} 2nd. Ed. (Springer, 2003)
\bibitem{arny86}V. I. Arnold, \emph{Catastrophe theory,} 2nd Edition (Springer, 1986)
\bibitem{arny92}V. I. Arnold, \emph{Ordinary differential equations}, 2nd Edition (Springer, 1992)
\bibitem{ap92}D. K. Arrowsmith and C. M. Place, \emph{Dynamical Systems} (Chapman and Hall, 1992)
\bibitem{tao}T. Tao, \emph{Nonlinear Dispersive Equations: Local and Global Analysis} (AMS, 2006)
\bibitem{carr81}J. Carr, \emph{Applications of center manifold theory} (Springer, 1981)
\bibitem{gg73}M. Golubitsky and V. Guillemin, \emph{Stable Mappings and Their Singularities }(Springer, 1973)
\bibitem{arny85} V.I. Arnold, S.M. Gusein-Zade, A.N. Varchenko, Singularities of Differentiable Maps, The classification of critical points, caustics, and wave fronts, Volume 1 (Springer, 1985)
\bibitem{arny2012}V.I. Arnold, S.M. Gusein-Zade, A.N. Varchenko, \emph{Singularities of Differentiable Maps, Monodromy and Asymptotics of Integrals}, Volume 2 (Springer, 2012)
\bibitem{gs85}M. Golubitsky and D. G. Schaeffer, \emph{Singularities and Groups in Bifurcation Theory,
Volume I} (Springer, 1985)
\bibitem{gss}M. Golubitsky, I, Stewart, and D. G. Schaeffer, \emph{Singularities and Groups in Bifurcation Theory,
Volume II} (Springer, 1988)
\bibitem{perko}L. Perko, \emph{Differential equations and dynamical systems}, 3rd Ed. (Springer, 2001)
\bibitem{arny88}V. I. Arnold and Yu. S. Il'yashenko,  \emph{ Ordinary differential equations,}  In: \emph{Dynamical Systems I}, D. V. Anosov and V. I. Annold,  (Eds.) (Springer, 1988)
\bibitem{sw}R. K. Sachs and H. Wu, \emph{General relativity for mathematicians} (Springer, 1977)

\end{thebibliography}
\end{document}